\UseRawInputEncoding
\documentclass[twocolumn]{aastex62}

\received{ }
\revised{ }
\accepted{ }
\submitjournal{ApJ}

\shorttitle{The variable jet of SAX J1808.4-3658}
\shortauthors{Baglio M. C. et al.}

\usepackage{multirow}
\begin{document}

\title{Probing jet launching in neutron star X-ray binaries: the variable and polarized jet of SAX J1808.4-3658}

\correspondingauthor{Maria Cristina Baglio}
\email{mcb19@nyu.edu}

\author[0000-0003-1285-4057]{M. C. Baglio}
\affiliation{Center for Astro, Particle and Planetary Physics, New York University Abu Dhabi, PO Box 129188, Abu Dhabi, UAE \\}
\affiliation{INAF, Osservatorio Astronomico di Brera, Via E. Bianchi 46, I-23807 Merate (LC), Italy}

\author[0000-0002-3500-631X]{D. M. Russell}
\affiliation{Center for Astro, Particle and Planetary Physics, New York University Abu Dhabi, PO Box 129188, Abu Dhabi, UAE \\}

\author[0000-0002-7441-5300]{S. Crespi}
\affiliation{Center for Astro, Particle and Planetary Physics, New York University Abu Dhabi, PO Box 129188, Abu Dhabi, UAE \\}

\author{S. Covino}
\affiliation{INAF, Osservatorio Astronomico di Brera, Via E. Bianchi 46, I-23807 Merate (LC), Italy}

\author{A. Johar}
\affiliation{Center for Astro, Particle and Planetary Physics, New York University Abu Dhabi, PO Box 129188, Abu Dhabi, UAE \\}

\author{J. Homan}
\affiliation{Eureka Scientific, Inc., 2452 Delmer Street, Oakland, CA 94602, USA\\}
\affiliation{SRON Netherlands Institute for Space Research, Sorbonnelaan 2, 3584 CA, Utrecht, The Netherlands\\}

\author{D. M. Bramich}
\affiliation{Center for Astro, Particle and Planetary Physics, New York University Abu Dhabi, PO Box 129188, Abu Dhabi, UAE \\}

\author{P. Saikia}
\affiliation{Center for Astro, Particle and Planetary Physics, New York University Abu Dhabi, PO Box 129188, Abu Dhabi, UAE \\}

\author[0000-0001-6278-1576]{S. Campana}
\affiliation{INAF, Osservatorio Astronomico di Brera, Via E. Bianchi 46, I-23807 Merate (LC), Italy}

\author{P. D'Avanzo}
\affiliation{INAF, Osservatorio Astronomico di Brera, Via E. Bianchi 46, I-23807 Merate (LC), Italy}

\author{R. P. Fender}
\affiliation{University of Oxford, Department of Physics, Astrophysics, Denys Wilkinson Building, Keble Road, OX1 3RH, Oxford, United Kingdom}

\author{P. Goldoni}
\affiliation{APC, Astroparticule et Cosmologie, Universit\'{e} Paris Diderot, CNRS/IN2P3, CEA/Irfu, Observatoire de Paris, Sorbonne Paris Cit\'{e}, 10, Rue Alice Domon et L\'{e}onie Duquet, F-75006 Paris, France}

\author{A. J. Goodwin}
\affiliation{School of Physics and Astronomy, Monash University, Clayton, 3800, Australia}

\author[0000-0003-3352-2334]{F. Lewis}
\affiliation{Faulkes Telescope Project, School of Physics and Astronomy, Cardiff University, The Parade, Cardiff, CF24 3AA, Wales, UK}
\affiliation{Astrophysics Research Institute, Liverpool John Moores University, 146 Brownlow Hill, Liverpool L3 5RF, UK}

\author[0000-0001-9487-7740]{N. Masetti}
\affiliation{INAF Ð Osservatorio di Astrofisica e Scienza dello Spazio, via Piero Gobetti 93/3, I-40129 Bologna, Italy}
\affiliation{Departamento de Ciencias F\`{i}sicas, Universidad Andr\'{e}s Bello, Av. Fern\'{a}ndez Concha 700, 7591538 Las Condes, Santiago, Chile}

\author[0000-0002-0943-4484]{A. Miraval Zanon}
\affiliation{Universit\`a dell'Insubria, Dipartimento di Scienza e Alta Tecnologia, Via Valleggio 11, I-22100, Como, Italy}
\affiliation{INAF, Osservatorio Astronomico di Brera, Via E. Bianchi 46, I-23807 Merate (LC), Italy}

\author{S. E. Motta}
\affiliation{INAF, Osservatorio Astronomico di Brera, Via E. Bianchi 46, I-23807 Merate (LC), Italy}
\affiliation{University of Oxford, Department of Physics, Astrophysics, Denys Wilkinson Building, Keble Road, OX1 3RH, Oxford, United Kingdom}

\author[0000-0002-3348-4035]{T.~Mu\~noz-Darias}
\affiliation{Instituto de Astrof\'{i}sica de Canarias, 38205 La Laguna, Tenerife, Spain}
\affiliation{Departamento de Astrof\'{i}sica, Universidad de La Laguna, E-38206 La Laguna, Tenerife, Spain}

\author{T. Shahbaz}
\affiliation{Instituto de Astrof\'{i}sica de Canarias, 38205 La Laguna, Tenerife, Spain}
\affiliation{Departamento de Astrof\'{i}sica, Universidad de La Laguna, E-38206 La Laguna, Tenerife, Spain}

\begin{abstract}
We report on an optical photometric and polarimetric campaign on the accreting millisecond X-ray pulsar (AMXP) SAX J1808.4-3658 during its 2019 outburst. 
The emergence of a low-frequency excess in the spectral energy distribution in the form of a red excess above the disc spectrum (seen most prominently in $z$, $i$ and $R$-bands) is observed as the outburst evolves. This is indicative of optically thin synchrotron emission due to a jet, as seen previously in this source and in other AMXPs during outburst.
At the end of the outburst decay, the source entered a reflaring state. The low-frequency excess is still observed during the reflares. Our optical ($BVRI$) polarimetric campaign shows variable linear polarization (LP) throughout the outburst. We show that this is intrinsic to the source, with low-level but significant detections (0.2--2$\%$) in all bands. The LP spectrum is red during both the main outburst and the reflaring state, favoring a jet origin for this variable polarization over other interpretations, such as Thomson scattering with free electrons from the disc or the propelled matter. During the reflaring state, a few episodes with stronger LP level (1--2 \%) are observed. The low-level, variable LP is suggestive of strongly tangled magnetic fields near the base of the jet. These results clearly demonstrate how polarimetry is a powerful tool for probing the magnetic field structure in X-ray binary jets, similar to AGN jets.
\end{abstract}



\section{Introduction} \label{sec:intro}
Millisecond pulsars (MSPs) are low--magnetic field ($\sim10^8$ G) neutron stars (NSs) emitting pulsed radiation with a periodicity of 1--10 ms (typically in the radio band, but also in the X- and $\gamma$-rays). The most accredited scenario able to explain the existence of such objects is the so called recycling scenario of MSPs (e.g. \citealt{Srinivasan2010}). According to this, MSPs would be NSs that have been accelerated or recycled (i.e. switched on again with lower magnetic field and higher rotational period) through accretion, typically in low mass X-ray binaries (LMXBs; \citealt{Alpar82}; \citealt{Radhakrishnan82}).
After decades of search, the first LMXB pulsating in the X-rays at the ms period (accreting millisecond X--ray pulsar -- AMXP) was discovered in 1998 (SAX J1808.4-3658; \citealt{Wijnands}), with 2.49 ms X-ray pulsations, showing that NSs in LMXBs can indeed spin very rapidly. This conclusion was then reinforced with the detection of coherent X--ray pulsations in the ms regime in other LMXBs. Twenty-two AMXPs have been discovered so far (for reviews see \citealt{PatrunoReview,Campana2018}). 

Optical and near infrared (NIR) light emitted from AMXPs (and NS-LMXBs in general) is traditionally thought to originate in the reprocessing of the X-ray thermal emission from the accretion disc; this process is dominant during outbursts in transient systems, whereas during quiescence the major contribution at these wavelengths is given by the low-mass companion star. In addition, evidence for the emission of relativistic jets has been found at optical-NIR (OIR) frequencies. \citet{Russell2007} showed that reprocessing of X-ray thermal emission from accretion discs cannot explain the observed OIR fluxes in Atoll sources and AMXPs in their high-luminosity states (as well as for black hole LMXBs in hard X-ray states). In particular, optically thin synchrotron emission from jets is found to dominate the NIR light when $L_{X} >10^{36} \, \rm erg \, \rm s^{-1}$, and the optical for $L_{X} >10^{37} \, \rm erg \, \rm s^{-1}$ \citep{Russell2007}. For NS LMXB Z-sources, X-ray reprocessing can instead explain well the observed OIR fluxes, with synchrotron emission possibly contributing to the NIR in most cases \citep{Russell2007}.

Evidence for the production of relativistic jets in AMXPs can be found from radio detections \citep[e.g.][and references therein]{Gaensler1999,Tudor2017}, and OIR spectral energy distributions (SEDs). If a jet is produced, a flux excess at the lowest OIR frequencies can be observed, due to the superposition of the optically thin synchrotron jet spectrum and the low-frequency tail of the blackbody of the reprocessed accretion disc. This excess has been observed in four AMXPs so far \citep{Russell2007}: XTE J1814--338 (\citealt{Krauss2005}) during outburst \citep[evidence of a quiescent residual jet is also reported in][]{Baglio2013}, SAX J1808.4--3658 \citep{Wang2001,Greenhill2006,Baglio2015ATel7469}, IGR J00291+5934 \citep{Torres2008,Lewis2010} and XTE J0929--314 \citep{Giles2005}. The NIR excess has also been detected in other NS-LMXBs (e.g. \citealt{Callanan2002}, \citealt{Migliari10}, \citealt{Harrison2011}, \citealt{Harrison2014}, \citealt{Wang2014}, \citealt{Baglio2014b}, \citealt{Baglio2016a}, \citealt{Baglio2019PSRJ1023}; see \citealt{Saikia2019} and references therein for examples of NIR excesses in black hole LMXBs). At lower frequencies, the spectrum of the jet is optically thick, characterized by a flat or slightly inverted radio slope. The break between the optically thick and thin spectrum typically falls in the NIR (jet break frequency; \citealt{Falcke2004}). The break has been detected in the NS LMXBs 4U 0614+091 and Aql X--1 \citep{Migliari10,DiazTrigo2018} and in the BH LMXBs GX 339-4 (\citealt{Corbel02}, \citealt{Gandhi11}, \citealt{CadolleBell2011}, \citealt{Corbel2013}), XTE J1118+480 \citep{Hynes2006}, V404 Cyg \citep{Russell2013a,Tetarenko2019}, XTE J1550-564 \citep{Chaty2011}, MAXI J1836-194 \citep{Russell2013b} and MAXI J1535-571 (Russell et al. 2020, submitted).

Synchrotron emission is one of the physical processes capable of generating linearly polarized radiation, up to very high levels \citep{Ribicky79}. Thus the case is strong that there can be a significant presence of linearly polarized emission from LMXBs that hints for the presence of jets. In particular, different behaviours of the linear polarization (LP) are expected, depending on the frequency of the observation. If the frequency is in the optically thick jet spectrum regime, a maximum of a few percent LP is generally observed \citep[e.g.][]{Fender01}. Optically thin synchrotron could instead produce LP up to $\sim 70\%$ \citep{Blandford02}; however, in NIR observations (i.e. close to the jet-break frequency), this typically translates into a significantly lower value due to the low ordering of the magnetic field lines in the jet (as in the case of the NS LMXB 4U 0614+091; \citealt{Baglio2014b}). Nevertheless, some LMXBs showing up to $30\%$ LP have been found, indicating that a high level of magnetic field lines ordering for LMXBs is also possible (\citealt{Fender01}; see the case of the BH LMXB Cyg X-1 reported in \citealt{Russell14}). Polarimetry has therefore the power to probe the magnetic field structure in jets.

At optical frequencies instead, the expected LP should never exceed a few per cent, due to the dominant emission coming from non-polarized sources at those frequencies, such as the accretion disc or the companion star \citep{Shahbaz08}. See \citet{Shahbaz2019Review} for a recent review on polarimetry of LMXBs. 
OIR LP can also arise from different processes, like electronic (Thomson) scattering of radiation with free electrons that can be found, e.g, in the geometrically symmetrical accretion disc, that is mostly ionised. Since each electron that is responsible for the scattering in the disc oscillates in principle in a random direction, a very low and constant level of LP is expected (a few per cent at the maximum), and the LP will be higher for higher frequencies, following the spectrum of the accretion disc. Sometimes, a significant phase-dependent variability of the polarization with the orbital period can be observed (see e.g. the cases of GRO J1655--40, Her X--1 and PSR J1023+0038; \citealt{Gliozzi98}; \citealt{Schultz04}; \citealt{Baglio2016b}).

\section{SAX J1808.4--3658} \label{sec:style}
The X-ray binary SAX J1808.4--3658 (hereafter SAX J1808) is an X-ray transient discovered in 1996 \citep{Zand1998} at a distance of 2.5--3.5 kpc (\citealt{Zand1998}; \citealt{Galloway2006}). It is the first AMXP discovered \citep{Wijnands}, the compact object being a 401 Hz ($\sim$ 2.5 ms) X-ray pulsar. The companion star has been proposed to be a semi-degenerate star with a mass of $\sim$ 0.05 - 0.10 $M_{\odot}$ (\citealt{Bildsten2001}; \citealt{Wang2001}; \citealt{Deloye2008}). The system has an orbital period of $\sim$2 hr \citep{Chakrabarty98} and the NS magnetic field has been inferred to be $\sim 10^8$ G \citep{Psaltis99}.

SAX J1808 displays recurrent outbursts. To date, nine outbursts have been observed (in 1996, 1998, 2000, 2002, 2005, 2008, 2011, 2015, 2019), with a recurrence time of 2-4 years. These events typically show a main outburst with a fast rise and a subsequent slow decay. After this phase, which lasts for approximately one month, a reflaring period at low X-ray luminosities ($10^{32-35}$ erg/s) is seen, which persists for tens of days (see e.g., \citealt{Wijnands2001}; \citealt{Patruno2016_1808}). This reflaring state has been explained by invoking a propeller scenario, with a large amount of matter expelled from the system, or the presence of a trapped disc truncated at the co-rotation radius \citep{Patruno2016_1808}.

Optical and infrared photometry of the system during its 1998 outburst revealed the presence of a clear infrared excess with respect to the expected low-frequency blackbody tail of the accretion disc and/or companion star \citep{Wang2001}, which was again seen in the 2005 outburst \citep{Greenhill2006}. This behaviour is similar to what has been observed in the case of other AMXPs (e.g. XTE J1814--338, IGR J00291+5934, XTE J0929--314; e.g. \citealt{Krauss2005}, \citealt{Giles2005}, \citealt{Lewis2010}), and has been indirectly linked to synchrotron emission from jets during outburst (signatures of a quiescent, receding jet have also been reported for XTE J1814--338 in \citealt{Baglio2013}).

During the 2015 outburst, an infrared ($H$-band) excess was reported in \citet{Baglio2015ATel7469}. Additionally, NIR polarimetry was obtained with the SofI instrument mounted on the New Technology Telescope (NTT, La Silla, Chile; ID: 295.D-5012, PI: Baglio), with the aim of confirming the synchrotron nature of the NIR emission. See Sec. \ref{Sec_2015} for further details on this.

SAX J1808 has been monitored at optical wavelengths since 2008 with the Las Cumbres Observatory (LCO) network of robotic 1-m and 2-m telescopes, including the 2-m Faulkes Telescope South at Siding Spring, Australia \citep{Elebert2009,Tudor2017}. As part of this campaign, after several years of quiescence, between 2019 July 25th and 30th a brightening was observed \citep{Russell2019ATel12964}, and interpreted as the beginning stages of a new outburst, or an optical precursor (see e.g. \citealt{Bernardini2016}, \citealt{Russell2019}).
The brightening was not observed in the X-rays until August 6th with \textit{Swift}/XRT
\citep{Goodwin2019ATel12993} and NICER \citep{Goodwin2020}. According to NICER observations, the outburst reached its peak on August 13th, after which it started to decay \citep{Bult2019ATel13077}. On August 24th the source entered its reflaring state, starting to alternate between lower ($\sim 10^{32}$ erg/s) and higher ($\sim 10^{33}$ erg/s) luminosity periods on $\sim$day timescales \citep{Bult2019ATel13077}. The reflaring state was also observed at optical frequencies thanks to the Faulkes/LCO on-going monitoring \citep{Baglio2019ATel13103}. At the beginning of October 2019, optical observations revealed that the source was back to its quiescent level \citep{Baglio2019ATel13162}.

\section{Observations and data analysis}
SAX J1808 was extensively monitored during its 2019 outburst using the Las Cumbres Observatory (LCO) 2\,m (Faulkes) and 1\,m telescopes in the $B, V, R, i'$ optical bands, and with the Very Large Telescope (VLT) in Cerro Paranal, Chile, equipped with the FORS2 instrument used in polarimetric mode ($B, V, R, I$ bands; (Program ID:0103.D-0575; PI: Baglio).

\subsection{VLT polarimetry}\label{polla_description}
We observed SAX J1808 with VLT/FORS2 in polarimetric mode with the optical filters $I\_BESS$$+77$, $R\_SPECIAL$$+76$, $v\_HIGH$$+114$, $b\_HIGH$$+113$ ($I$, $R$, $V$, $B$), for 9 epochs between August 8th, 2019 and September 24th, 2019, covering approximately the entire outburst. Most observations were performed under optimal sky transparency conditions, which are required in order to obtain a good signal-to-noise in the polarimetric images.

\begin{table}[htb]
\caption{Detailed log of the VLT/FORS2 polarimetric observations performed between Aug. 8, 2019 and Sep. 24, 2019. In the last column, the sky transparency is reported. From the ESO website, clear conditions (CLR) imply less than 10$\%$ of the sky (above 30 degrees elevation) covered in clouds, and transparency variations under 10$\%$; thin conditions (THN) imply transparency variations above 10$\%$; thick conditions instead mean that large transparency variations are possible, that is equivalent to not having constraints on the transparency conditions\footnote{\url{https://www.eso.org/sci/observing/phase2/ObsConditions.VISIR.html}  }. }
\label{log_polla}      
\centering                       
\begin{tabular}{c c c c c}       
\hline             
Epoch (UT) & MJD        &  Filter & Exp. & Sky \\   
 (Label) & (start)    &          & time (s)       & transp.         \\
\hline                     
2019-08-08 & 58703.2102332 & b & 80  & CLR \\
 (Epoch 1)&  & v & 40  & CLR \\
\hline
2019-08-13 & 58708.2469446 & b & 80 & THN \\
 \multirow{3}{*}{(Epoch 2)}&  & v& 40  & THN \\
 &  & R & 40  & THN \\
&  & I & 60x2  & THN \\
\hline
2019-08-16 & 58711.1873337 & b & 80 & CLR \\
 \multirow{3}{*}{(Epoch 3)}&  & v& 40  & CLR \\
 &  & R & 40  & CLR \\
&  & I & 60x2  & CLR \\
\hline
2019-08-18 & 58713.1245152 & b & 80 & THICK \\
\multirow{3}{*}{(Epoch 4)} &  & v& 60  & THICK \\
 &  & R & 60  & THICK \\
&  & I & 80x2  & THICK \\
\hline
2019-08-21 & 58716.1522254 & b & 80 & CLR \\
\multirow{3}{*}{(Epoch 5)} &  & v& 60  & CLR \\
 &  & R & 60  & CLR \\
&  & I & 80x2  & CLR \\
\hline
2019-08-24 & 58719.9839237 & b & 80 & CLR \\
\multirow{3}{*}{(Epoch 6)} &  & v& 80  & CLR \\
 &  & R & 80  & CLR \\
&  & I & 100x2  & CLR \\
\hline
2019-09-04 & 58730.9961515 & b & 120 & THN\\
\multirow{3}{*}{(Epoch 7)} &  & v& 100  & THN \\
 &  & R & 70  & THN \\
&  & I & 130x2  & THN \\
\hline
2019-09-07 & 58733.0045035 & b & 150 & CLR \\
\multirow{3}{*}{(Epoch 8)} &  & v& 100  & CLR \\
 &  & R & 110  & CLR \\
&  & I & 180x2  & CLR \\
\hline
2019-09-25 & 58751.0161424 & b & 150  & CLR \\
(Epoch 9) &  & v & 100  & CLR \\

\hline                             
\end{tabular}
\end{table}

A Wollaston prism was inserted in the optical path, which allowed the incident radiation to be split into two simultaneous beams characterised by orthogonal polarization. The two beams are commonly referred to as ordinary (o-) and extraordinary (e-). To avoid overlapping of the beams on the CCD, a Wollaston mask was also introduced. Finally, the instrument made use of a rotating half wave plate (HWP) to take images at four different angles $\Phi_i$ with respect to the telescope axis: $\Phi_{i}=22.5^{\circ}(i-1), \, i=1, 2, 3, 4$. A log of the observations is reported in Table \ref{log_polla}. All images were reduced by subtracting an average bias frame and dividing by a normalized flat field. Aperture photometry was then performed using the {\tt daophot} tool \citep{Stetson2000}. 
The normalised Stokes parameters for the LP, $Q$ and $U$, are defined as follows:

\begin{equation}\label{Q_U_eq}
Q= \frac{F(\Phi_1)-F(\Phi_3)}{2}; \,\,\, U=\frac{F(\Phi_2)-F(\Phi_4)}{2} ,
\end{equation}

where:

\begin{equation}
F(\Phi_i)=\frac{f^o(\Phi_i)-f^e(\Phi_i)}{f^o(\Phi_i)+f^e(\Phi_i)},
\end{equation}

with $f^o$ and $f^e$ being the ordinary and extraordinary beams fluxes, respectively.
These parameters then need to be corrected for the instrumental polarization, which can be estimated thanks to the observation of a non-polarized standard star\footnote{\url{ https://www.eso.org/sci/facilities/paranal/instruments/fors/tools/FORS1_Std.html}}. For FORS2, unpolarized standard stars are frequently observed in order to monitor the level of the instrumental polarization\footnote{\url{ http://www.eso.org/observing/dfo/quality/FORS2/reports/FULL/trend_report_PMOS_inst_pol_FULL.html}}, which has been stable during the last 10 years in all bands, at a very low level ($0-0.3\%$).

Using these corrected Stokes parameters, one could evaluate the level of LP ($P$) and the angle of polarization (PA) $\theta$ as $P=(Q^2+U^2)^{0.5}$ and $\theta = (1/2)tan^{-1}(U/Q)$. The PA then needs to be corrected for instrumental effects that can be estimated thanks to the observation of a highly polarized standard star, whose PA is known and tabulated.

However, $P$ is not distributed as a Gaussian \citep{Wardle1974}. In order to take this into account, as shown in \citet{Wardle1974}, it would be necessary to add a bias correction to the measured polarization, which can become very large in case of low polarization and large errors, that is typically the case for LMXBs.

\subsubsection{The S-parameter method}

A very accurate algorithm has been developed in order to estimate the LP of a source, accounting for all the factors mentioned above, such as the non-Gaussian distribution of $P$ and the possible interstellar contribution to the measured LP \citep[e.g.][and references therein; see also \citealt{Covino1999}, \citealt{Baglio_cen2014}]{Serego}. This algorithm consists of the evaluation of a parameter $S(\Phi)$ for each HWP angle $\Phi$, defined as follows:
\begin{equation}
S(\Phi)=\left( \frac{f^{o}(\Phi)/f^e(\Phi)}{f^o_u(\Phi)/f^e_u(\Phi)}-1\right)/\left( \frac{f^{o}(\Phi)/f^e(\Phi)}{f^o_u(\Phi)/f^e_u(\Phi)}+1\right),
\end{equation}
which combines the ordinary and extraordinary fluxes for each HWP angle $\Phi$ measured for the target ($f^o(\Phi)$ and $f^e(\Phi)$, respectively) with the same quantities measured for an unpolarised (u) field star ($f^o_u(\Phi)$, $f^e_u(\Phi)$).
This parameter can be considered as the component of the normalised Stokes vector which describes the LP along the direction $\Phi$ (for example, $S(\Phi=0)=Q$). This quantity is then related to the polarization $P$ of the target and to the polarization angle $\theta$ by the following relation:
\begin{equation}\label{eq_cos}
S(\Phi) = P\, \cos 2(\theta-\Phi).
\end{equation}

Therefore, from the fit of the $S$ parameter with Eq. \ref{eq_cos}, one can obtain an accurate estimate of the LP and PA for the target (the latter however will still need to be corrected using the measured angle of a polarized standard star). Moreover, this method can become even more accurate by considering more than one reference field star. In this way, if $n$ is the number of the reference field stars, we will end up with $n$ different $S$-values for each angle, which will allow us to derive the probability distribution of $S(\Phi)$, without approximating it with a Gaussian function. 
The only assumption for this algorithm is that the chosen field stars are considered to be intrinsically unpolarized, or to be polarized only by the interaction with the interstellar medium (that we can suppose to be almost the same for the field stars and for the target itself). We notice that the majority of the reference stars chosen for this work have a Gaia DR2 position, and they were found to be all quite close to Earth, with an average distance of $2.5\pm0.6$ kpc (i.e. very similar to the one inferred for SAX J1808; \citealt{Zand1998}). 
According to \citet{Serkowski75}, it is possible to evaluate the maximum interstellar contribution to the LP ($P_{\rm int}$) that we expect to observe for any target, depending on its absorption coefficient in the $V$-band ($A_V$). In particular, $P_{\rm int}<3A_{V}$. For SAX J1808, the hydrogen column density $N_{\rm H}$ was measured during the last outburst to be $(1.46 \pm 0.12)\times 10^{21}\, \rm cm^{-2} 
$ \citep{DiSalvo2019}; considering the relation between $N_{H}$ and $A_{V}$ for our Galaxy reported in \citet{Foight16}, we obtain a dust extinction of $A_V=0.51\pm0.04$ and, therefore, $P_{\rm int}<1.53\%$.
In addition, as already mentioned, the instrumental polarization for the FORS2 instrument is expected to be very low\footnote{\url{http://www.eso.org/observing/dfo/quality/FORS2/reports/FULL/trend_report_PMOS_inst_pol_FULL.html}} (in our campaign we also observed a non-polarized standard star in three different epochs, always obtaining $3\sigma$ upper-limits to $P$ of $\sim 0.5\%$ in all bands). 

\begin{figure}
\centering
\includegraphics[scale=0.35]{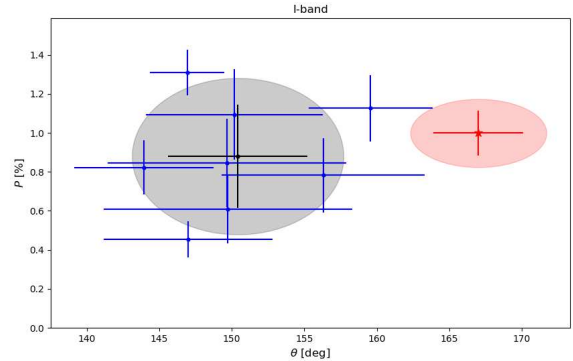}
\caption{Representation of $P$ vs. $\theta$ for the $I$-band observation of August 15th 2019. With blue dots, the reference stars are plotted (the black dot is the average of the reference stars). In red, SAX J1808 is shown. The error of the average is evaluated as the standard deviation of the N reference stars, divided by the square root of N, to account for the dispersion of the reference stars. Grey and red ellipsoids show the 68$\%$ confidence level regions for the average of the reference stars and for the target, respectively.
}
\label{P_theta}
\end{figure}

To verify that the chosen field stars are intrinsically unpolarized, we plotted them in the $P- \theta$ plane; ideally, the reference stars should all cluster around a common point in the plane with small dispersion, for each epoch and band of observation; any deviation of the centroid of their distribution from the origin of the plane is therefore a good estimate of the average instrumental and/or interstellar polarization along the line of sight. In Fig. \ref{P_theta} we show this graph for one band and one epoch ($I$ band on August 15th) for the target and for the group of reference stars. To build this figure, LP and PA values have been evaluated for each object starting from their Stokes parameters $Q$ and $U$ (eq. \ref{Q_U_eq}). No correction has been applied to the LP and PA values of any object at this stage. It is evident from the figure that the reference stars all cluster around a common value in the plane. The instrumental and interstellar LP estimated from this graph is low as expected ($<1\%$, consistently with what is obtained from \citealt{Serkowski75}). The PA of the group of reference stars is always of the order of $\sim 150^{\circ}$, in all epochs and bands.

In addition, we do not expect the polarization of the field stars to change from night to night. Therefore, the results of the analysis can become even more accurate by taking into account every epoch of all the reference stars on all nights, to correct the target values. The correction will be different from band to band. 
This is a good procedure because it allows any possible variability due to, for example, a different choice of reference stars from night to night, to average out, and provides more homogeneous results. 

Once $S(\Phi)$ is calculated, $P$ and $\theta$ are determined by maximizing the Gaussian likelihood function by an optimization algorithm \citep[e.g. the Nelder-Mead algorithm, ][]{Gao&Han2012} and integrating the posterior probability density of the parameters of our models by a Markov Chain Monte Carlo \citep[MCMC;][]{Hogg&Foreman2018} based on the ``affine-invariant Hamiltonian" algorithm \citep{Foreman-Mackeyetal2013}. We started the chains from small Gaussian balls centered on the best fit values. The first third of each chain (the ``burn-in phase") was discarded and we checked that a stationary distribution was reached \citep{Sharma2017}. Fit quality was evaluated as in \citet{Lucy2016}. The reported values for $P$ and $\theta$ and their 1\,$\sigma$ uncertainties are the $0.16, 0.50$ and $0.84$ quantiles of the posterior distribution of parameters. Therefore, in general, each $P$, $\theta$ pair taken together is not necessarily the best-fit solution although in most cases these values are very close to the maximum a posteriori estimates \citep[see][for an extensive discussion]{Hogg&Foreman2018}.


Finally, the value of $\theta$ obtained with this method was corrected by an additional factor from the observation of a polarized standard star, with known and tabulated polarization angle. The resulting average correction was very small in all bands and epochs ($<2^{\circ}$). 

\subsection{LCO monitoring}\label{lco_description}
We observed SAX J1808 with the Las Cumbres Observatory (LCO) 2\,m Faulkes and 1\,m network for the whole duration of the outburst, as part of a monitoring campaign of $\sim 50$ LMXBs (co-ordinated by the Faulkes Telescope Project\footnote{\url{http://www.faulkes-telescope.com/}}), which SAX J1808 was part of \citep{Lewis2008}. Imaging data were taken in the $B, R, V$ Bessell filters and the SDSS $u', i', z'$ filters (436.1--870 nm). Since the field is extremely crowded, the source was blended with a few nearby stars especially when SAX J1808 was faint (near and in quiescence), depending on the seeing conditions. In the majority of images however, the target could be disentangled and multi-aperture photometry (MAP; see \citealt{Stetson90}) was performed by the newly developed ``X-ray Binary New Early Warning System'' (XB-NEWS) pipeline \citep{Russell2019}. The pipeline automatically downloads new images of all the targets that are monitored with LCO, and all their associated calibration data; then, it performs several quality control steps, computes an astrometric solution for each image using Gaia DR2 positions\footnote{\url{https://www.cosmos.esa.int/web/gaia/dr2}}, performs MAP photometry (along with standard aperture photometry), solves zero-point calibrations between epochs as described in \citet{Bramich2012} and flux calibrates the photometry using the ATLAS-REFCAT2 catalogue (\citealt{Tonry2012}; which includes PanSTARRS\footnote{\url{ https://panstarrs.stsci.edu}} DR1, APASS\footnote{\url{https://www.aavso.org/apass}}, and other catalogues). The $R$-band is calibrated indirectly via comparison with predicted $R$-band standard magnitudes. The predicted $R$-band standard magnitudes are computed for ATLAS-REFCAT2 stars with Pan-STARRS1 $g_{\mbox{\scriptsize P1}}$ and $r_{\mbox{\scriptsize P1}}$ standard magnitudes using the transformations provided in \citet{Tonry2012}. The pipeline then produces a calibrated light curve for the target with near real-time measurements (for more details see \citealt{Russell2019,Pirbhoy2020}). In the cases for which the target was not automatically detected above the detection threshold by the pipeline, XB-NEWS performed forced multi-aperture photometry at the known position of the source. All magnitudes derived from forced photometry with an uncertainty $>0.25$ mag were considered unreliable, and were therefore rejected.  

Observations on similar dates as the VLT polarimetric data were taken in several filters.
A detailed observation log of the LCO data can be found in Table \ref{log_LCO}.

\begin{table*}[!htb]
\caption{Log of the LCO observations published in this work. Epochs included are those close in time to the VLT observations (Epochs 1-9) and two datasets in which timing observations were taken (MJD 58709 and 58722). 1m-A, 1m-C and 1m-S are 1-m nodes in Australia, Chile and South Africa, respectively.}            
\label{log_LCO}      
\centering                       
\begin{tabular}{c c c c c | c c c c c}       
\hline               
Epoch (UT) & MJD & Site & Filters & Exposure & Epoch (UT) & MJD & Site & Filters & Exposure \\   
\hline                     
2019-08-08                   & 58703.195069 & 1m-C  & $B$    & 200s & & 58719.876733 & 1m-C  & $B$    & 60s \\
(Epoch 1)           & 58703.183335 & 1m-C  & $V$    & 100s & &58719.869932 & 1m-C  & $V$    & 60s \\
                             & 58703.192862 & 1m-C & $R$  & 100s & (Epoch 6) & 58719.875553 & 1m-C & $R$  &  60s\\
                             & 58703.181707 & 1m-C  & $i'$   & 100s & 2019-08-24 & 58719.868769 & 1m-C  & $i'$   & 60s \\
                             & 58703.186128 & 1m-C  & $z'$   & 300s & & 58719.871904 & 1m-C  & $z'$   & 200s\\
\hline
2019-08-13                   & 58708.789980 & 1m-S  & $B$    & 30s & 2019-09-07 & 58732.748023 & 1m-S  & $i'$   & 200s \\
(Epoch 2)           & 58708.785300  & 1m-S  & $V$    & 30s & & & & & \\
                             & 58708.789043 & 1m-S & $R$  & 30s & & & & &\\
                             & 58708.784367 & 1m-S  & $i'$   & 30s & & & & & \\
                             & 58708.786590 & 1m-S  & $z'$   & 100s & & & & & \\
 
\hline
2019-08-14 & 58709.799704 & 1m-S & $i'$ & $20\times 100s$ & 2019-09-04 & 58731.545601 & 1m-C  & $B$    & 100s \\
& & & & & (Epoch 7)                                                                      & 58731.537881 & 1m-C  & $V$    & 100s \\
& & & & & &                                                                                         58731.535669 & 1m-C  & $i'$   & 200s \\
& & & & &                                                                                         & 58731.540106 & 1m-C  & $z'$   & 200s \\
\hline 
2019-08-16                   & 58710.836103 & 1m-S  & $B$    & 40s & 2019-09-07 & 58732.748023 & 1m-S  & $i'$   & 200s \\
(Epoch 3)           & 58710.831429 & 1m-S  & $V$    & 40s & (Epoch 8) & & & & \\
                             & 58710.835160 & 1m-S & $R$  & 40s & & & & & \\
                             & 58710.830488 & 1m-S  & $i'$   & 40s & & & & & \\
                             & 58710.832717 & 1m-S  & $z'$   & 100s & & & & & \\

\hline
2019-08-17                   & 58712.855179 & 1m-A  & $B$    & 40s & 2019-09-25          & 58751.849739 & 1m-S  & $B$    & 200s \\  
(Epoch 4)           & 58712.850531 & 1m-A  & $V$    & 40s & (Epoch 9)  & 58751.840343 & 1m-S  & $V$    & 100s\\
                             & 58712.854255 & 1m-A & $R$   & 40s &                       & 58751.847530 & 1m-S & $R$  & 100s\\
                             & 58712.849606 & 1m-A  & $i'$   & 40s &                     & 58751.838130 & 1m-S  & $i'$   & 200s\\
                             & 58712.851805 & 1m-A  & $z'$   & 100s &                    & 58751.842553 & 1m-S  & $z'$   & 200s \\
\hline 
2019-08-21                   & 58716.859265 & 1m-S  & $B$    & 60s & & & & &\\
 (Epoch 5)          & 58716.852543 & 1m-S  & $V$    & 60s & & & & &\\
                             & 58716.858100 & 1m-S & $R$  & 60s & & & & &\\
                             & 58716.851378 & 1m-S  & $i'$   & 60s & & & & &\\
                             & 58716.854512 & 1m-S  & $z'$   & 200s & & & & &\\
\hline
\hline                             
\end{tabular}
\end{table*}

\section{Results}\label{Sec:results}

\begin{figure*}[h]
\centering
\includegraphics[scale=0.64]{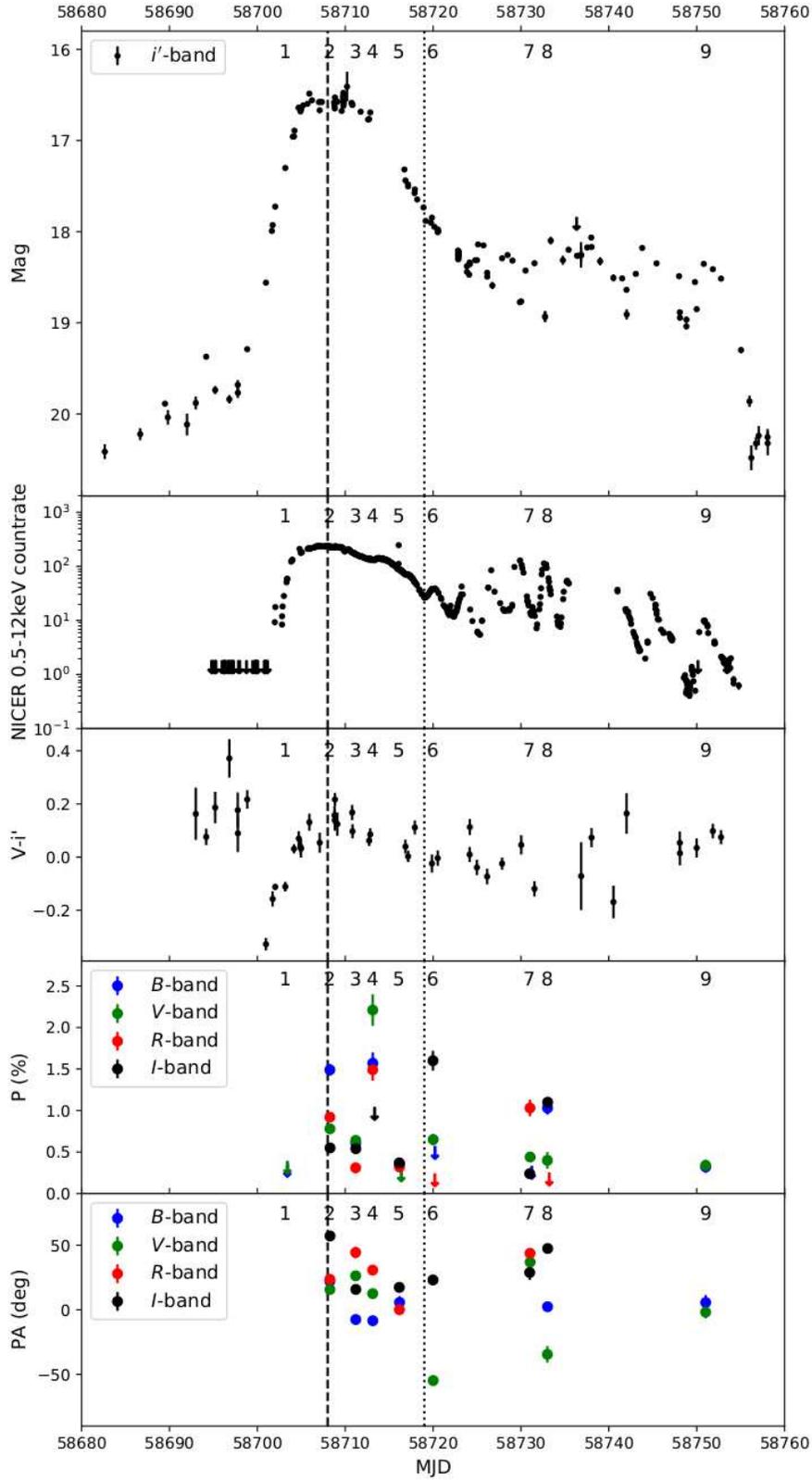}
\caption{From top to bottom, (1) LCO $i'$ band light curve of the outburst of SAX J1808; (2) NICER 0.5--10 keV light curve of the outburst; (3) absorbed $V-i'$ color trend with time during the outburst; (4) Polarization detections and upper limits in $B$, $V$, $R$ and $I$ bands (VLT/FORS2). With downward arrows, the $3\sigma$ upper limits are represented with the same colors as for the detections (dots); (5) Polarization angle vs. time.
In Plot (1) and (2), black arrows indicate 3$\sigma$ upper limits to the $i'$ band magnitude and the NICER count rate, respectively.
In all plots, a dashed vertical line indicates the time of the peak of the outburst (according to X-ray observations), whereas a dotted line shows the starting time of the X-ray reflaring state. With numbers (1-9), the epochs of the polarimetric observations are indicated in all panels, as defined in Tab. \ref{log_polla}.}
\label{fig_big}
\end{figure*}

\begin{figure*}
\centering
\includegraphics[scale=0.32]{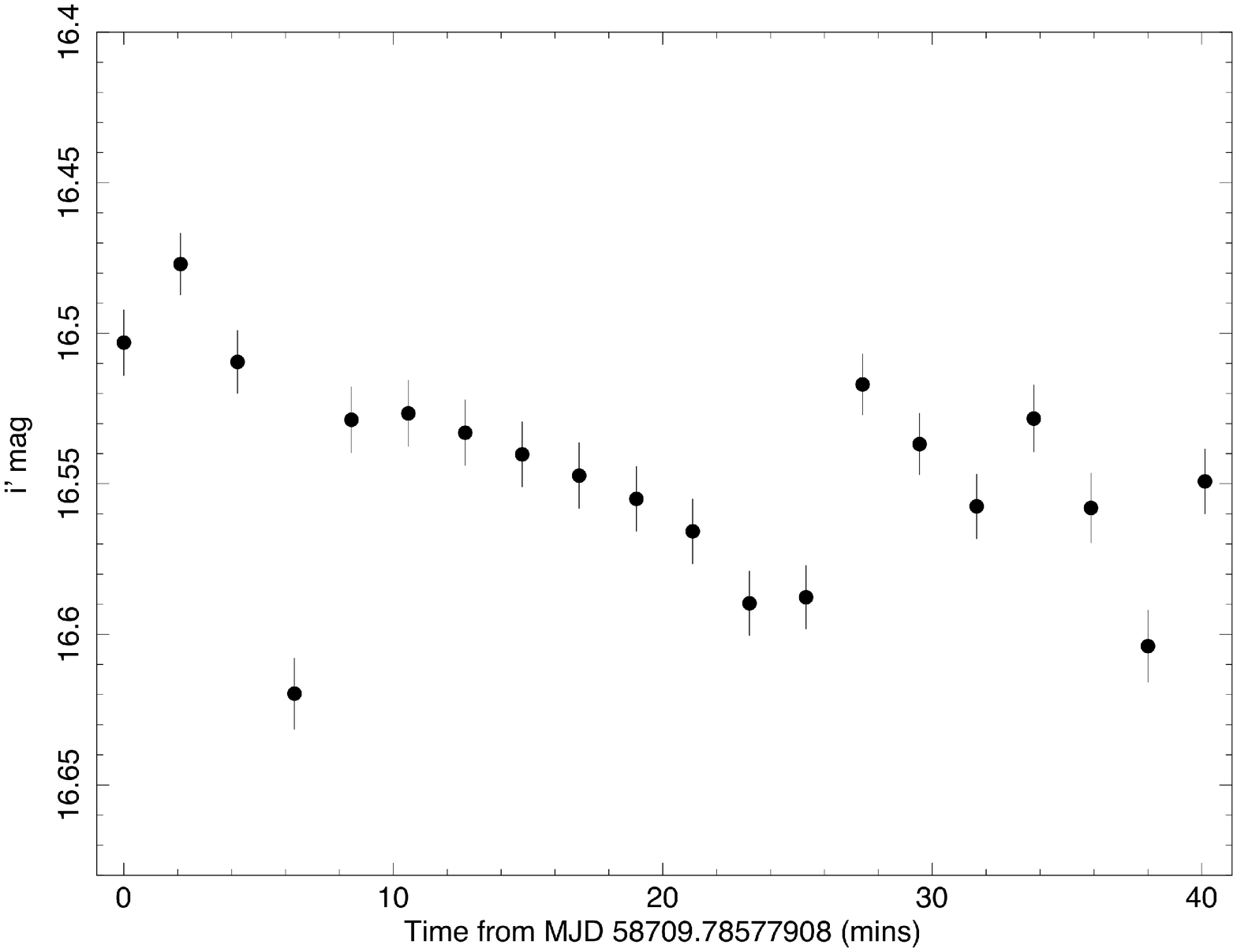}
\includegraphics[scale=0.32]{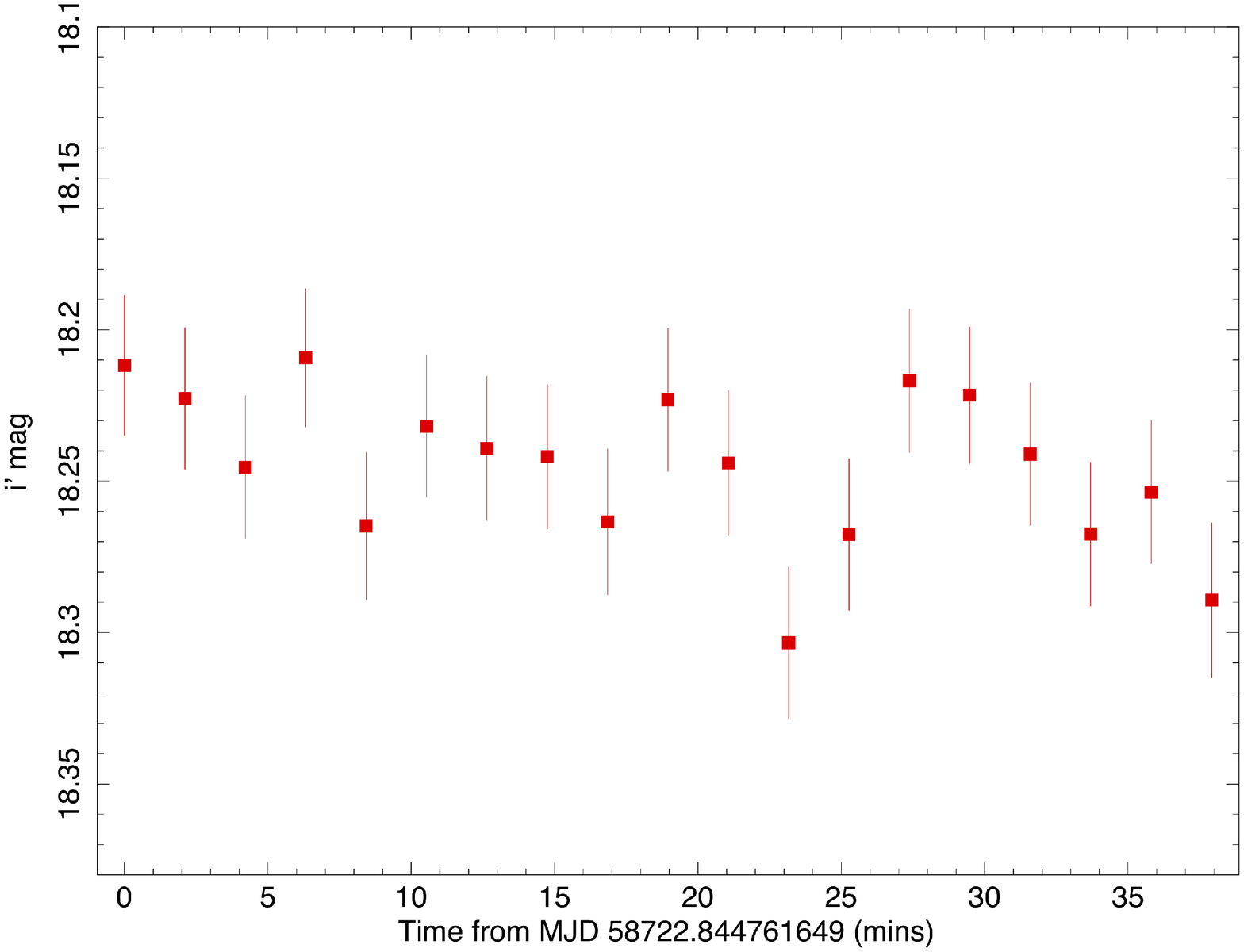}
\caption{$i'$ band light curves based on LCO data taken on MJD 58709 (2019 August 14; left panel) and 58722 (2019 August 27; right panel).}
\label{fast_timing_fig}
\end{figure*}

In Fig. \ref{fig_big} the main results of this campaign are shown. The $i'$ band light curve obtained with LCO (upper panel) shows a main outburst, that is also clearly observed in the NICER X-ray light curve (second panel; see \citealt{Bult2019}, \citealt{Goodwin2020} for details of the NICER campaign), followed by a reflaring state, with optical reflares of amplitude $\sim 0.5$ mag.
We note that prior to MJD 58701 (August 6th) all the NICER observations were background dominated; therefore, we took an upper limit of 1.37 counts/s, i.e. the average pre-outburst net count rate plus three times the 1-$\sigma$ standard deviation as the detection limit for SAX J1808 \citep{Goodwin2020}.
By comparing the optical to the X-ray light curve, we can see that the peak of the outburst was first reached in the optical at MJD $\sim 58706$ (August 11th), and then in X-rays at MJD 58709 (August 14th; see also \citealt{Goodwin2020}). A small optical brightening at almost the same time as the onset of the X-ray reflaring state is observed, possibly suggesting a similar starting time of the reflaring state in the two bands (this will be discussed in more details in a follow up paper; see Russell et al. 2020 in preparation).
In the third panel of Fig. \ref{fig_big} we report the evolution of the $V-i'$ absorbed colour during the outburst (evaluated using the LCO data). The $V-i'$ colour first decreases during the initial rise, then starts to increase during the fast rise of the outburst, meaning that the system is becoming first bluer, then redder. $V-i'$ reaches its peak value at the same time as the main outburst peaks, and then decreases with the luminosity, indicating that the spectrum of the source is becoming bluer during the fade from the main peak. Therefore, the behaviour of the $V-i'$ colour suggests the existence of a red component that brightens during the outburst rise, and then fades as the outburst decays. 
During the reflaring state, more scatter is observed in $V-i'$, hinting that there is a variable component, or suggesting the presence of short timescale variability which was not possible to sample for all epochs properly with the LCO monitoring campaign (see Fig. \ref{fast_timing_fig}). 

To investigate a possible optical short timescale variability, we looked at two $i'$-band fast timing (2 mins binning) light curves obtained using LCO data acquired on MJD 58709 (August 14th; at the peak of the outburst) and on MJD 58722 (August 27th; at the very beginning of the reflaring state; see Tab. \ref{log_LCO}). Both curves show signs of fast variability, with a measured fractional rms of $(3\pm1)\%$ on MJD 58709 over a 40-min timescale (see Fig. \ref{fast_timing_fig}). The fractional rms is instead not significant on MJD 58722. This hints at the presence of a rapidly (on tens of minutes timescales) variable component in the optical emission throughout the outburst.

The last two panels of Fig. \ref{fig_big} show the results of the VLT polarimetric campaign (see also Table \ref{log_polla}). The analysis of the data (performed as described in Sec. \ref{polla_description}) allowed us to reach very high precision in the determination of the fractional LP in case of a detection, with errors of $\sim 5-20 \%$ of the polarization level, depending on the night (see Table \ref{log_polla}). The results are reported in Table \ref{polla_results}.

\begin{table*}[!h]
\caption{Results of the VLT/FORS2 ($BVRI$ filters) polarimetric campaign on the 2019 outburst of SAX J1808. All the polarization levels and angles are corrected for instrumental polarization. The interstellar polarization has also been subtracted, by means of a group of reference stars in the field.  All the upper limits are indicated at a $3\sigma$ confidence level and have been evaluated as described in Sec. \ref{polla_description}.}            
\label{polla_results}      
\centering                       
\begin{tabular}{c |c |c |c |c |c| c |c |c}       
\hline               
 \multirow{2}{*}{Epoch} &\multicolumn{2}{c|}{$B$} & \multicolumn{2}{c|}{$V$} & \multicolumn{2}{c|}{$R$} & \multicolumn{2}{c}{$I$} \\   
 
 &$P(\%)$ & $\theta$($^{\circ}$)& $P(\%)$ & $\theta$($^{\circ}$) & $P(\%)$ & $\theta$($^{\circ}$) & $P(\%)$ & $\theta$($^{\circ}$)\\
\hline                       
1 &$< 0.27$ & -- & $< 0.31$ & -- & -- & -- & -- & -- \\
\hline
2 & $1.49\pm0.07$ & $22.13^{+1.32}_{-1.34}$ & $0.78^{+0.05}_{-0.06}$ &$15.71^{+2.03}_{-2}$ & $0.92\pm 0.05$ & $23.61 ^{+1.67}_{-1.68}$ & $0.55^{+0.05}_{-0.04}$ & $57.17^{+2.27}_{-2.23}$ \\
\hline
3& $0.61^{+0.08}_{-0.07}$ & $172.49^{+3.42}_{-3.36}$ & $0.64^{+0.06}_{-0.05}$ & $26.34 \pm 2.32$ & $0.31\pm0.05$ & $44.44^{+4.42}_{-4.41}$ & $0.54 ^{+0.04}_{-0.05}$ & $15.78^{+2.23}_{-2.25}$ \\
\hline
4 & $1.57^{+0.14}_{-0.13}$ & $171.57^{+1.72}_{-1.9}$ & $2.21 \pm 0.19$ & $12.55^{+2.20}_{-2.06}$ & $1.49^{+0.12}_{-0.13}$ & $30.75^{+2.80}_{-2.56}$ & $<0.96$ & -- \\
\hline
5 & $0.35\pm 0.08$ & $5.76^{+4.82}_{-4.86}$ & $<0.22$ & -- & $0.32^{+0.04}_{-0.05}$ & $0.15^{+4.29}_{-4.31}$ & $0.37^{+0.05}_{-0.04}$ &$17.39^{+3.32}_{-3.28}$ \\
\hline
6 & $< 0.49$ & -- & $0.65\pm0.07$ & $125.23^{+2.89}_{-3.74}$ & $<0.16$ & -- & $1.60^{+0.11}_{-0.12}$ &$23.12^{+1.93}_{-1.96}$ \\
\hline
7 &$< 0.25$ & -- & $0.44\pm0.06$ & $36.95^{+3.87}_{-3.79}$ & $<0.17$ & -- & $0.24\pm 0.05$ &$28.87^{+6.09}_{-5.82}$ \\
\hline
8 & $1.03^{+0.07}_{-0.08}$ & $33.32^{+2.39}_{-2.42}$ & $0.4^{+0.09}_{-0.10}$ & $145.53^{+6.25}_{-6.46}$ & $1.03^{+0.11}_{-0.10}$ & $43.73^{+2.86}_{-2.79}$ & $1.10^{+0.04}_{-0.05}$ &$47.47^{+1.28}_{-1.30}$ \\
\hline
9 & $0.32\pm0.07$ & $49.42^{+5.70}_{-5.66}$ & $0.34^{+0.06}_{-0.07}$ & $178.24^{+1.29}_{-4.8}$ & -- & -- & -- &-- \\

\hline

\end{tabular}
\end{table*}

We obtained several significant detections of LP in all bands. In case of non-significant detections, we evaluated $3\sigma$ upper limits in all bands, that are also reported in Table \ref{polla_results}, and are all quite constraining ($<1\%$), except for the few epochs in which the signal-to-noise was too low, due to bad seeing or thick sky transparency.

\begin{figure}
\centering
\includegraphics[scale=0.34]{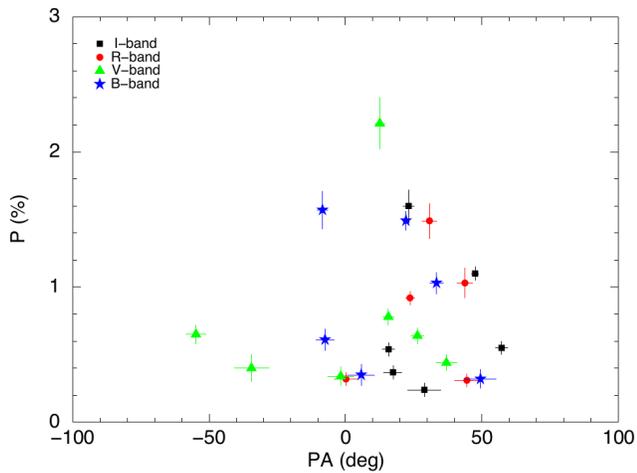}
\caption{Polarization level $P(\%)$ vs. Polarization angle PA$(^{\circ})$ in all epochs of the 2019 outburst for $I$ band (black squares), $R$ band (red circles), $V$ band (green triangles) and $B$ band (blue stars). }
\label{pol_angle_fig}
\end{figure}

For the epochs in which a detection of polarization is observed, we also evaluated the polarization angle PA. 
In Fig. \ref{pol_angle_fig} we show the polarization level $P$ vs. angle PA in all bands. The PA appears to be clustered around $\sim 0$--50$^{\circ}$, and no clear correlation between the two quantities can be found in any of the bands.


\subsection{The Moon contribution}\label{Sec:Moon}
\begin{figure*}
\centering
\includegraphics[scale=0.34]{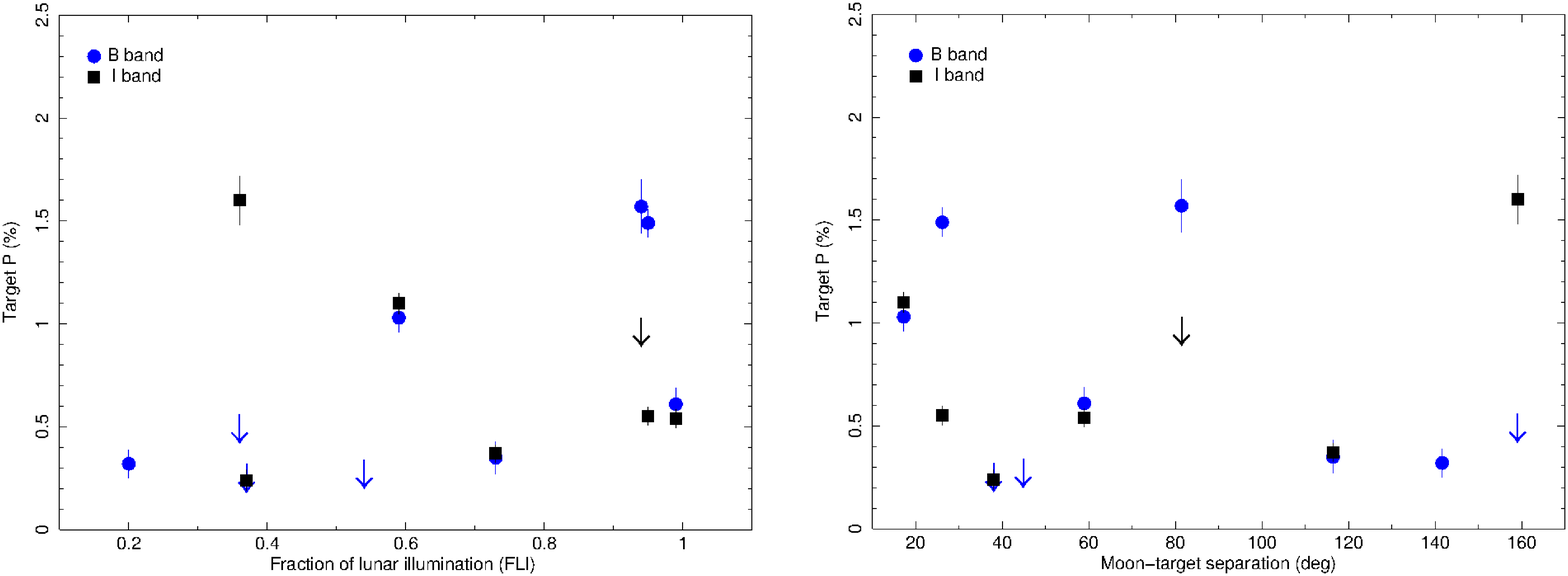}
\caption{\textit{Left panel}: $B$- (blue dots) and $I$- (black squares) band polarization of SAX J1808 vs. fraction of lunar illumination (FLI) during the 2019 outburst; \textit{Right panel}: $B$- (blue dots) and $I$- (black squares) band polarization of SAX J1808 vs. moon-target separation on the sky (deg). In both panels, with blue and black downwards arrows we indicate the $3\sigma$ upper limits of the $B$- and $I$-band polarization levels, respectively.}
\label{moon_fig}
\end{figure*}

Our polarimetric monitoring covered more than 30 days (Table \ref{log_polla}). Therefore, the phase of the Moon considerably changed, as well as its separation from the target. The presence of the Moon can alter the measured LP due to reflection, mostly in $B$ and $V$ bands. We therefore looked for a correlation between (i) the fraction of lunar illumination (FLI) and the measured level of polarization of the target $P$ $(\%)$ and (ii) the moon-target separation on the sky (deg) and $P$. In particular, the FLI is defined as the fraction of the lunar disc that is illuminated at local midnight, and it is equal to 1 when the disc is ``fully illuminated'', and 0 when the Moon is below the local horizon. 
The results of the analysis are shown in Fig. \ref{moon_fig} for the $B$-band and $I$-band. 

A hint of a correlation is present in both figure panels in $B$-band, with a level of LP that increases with increasing FLI (Pearson correlation coefficient $R$ for this correlation is 0.59) and decreasing Moon-target separation ($R\sim -0.64$). In particular, most upper limits are found at low FLI, i.e. when the effect of reflection should be at its minimum. 
In $I$-band instead, no correlation seems to be present ($R\sim 0.02$ for the Moon-target separation and $R\sim -0.40$ for the FLI). Similar results to the $I$-band ones are also obtained in $R$ and $V$ bands, with no correlation of the LP of SAX J1808 with the FLI of the Moon or with the Moon-target angular separation in the sky.

We therefore conclude that the $B$-band polarization detections reported in this work are affected by the presence of the Moon, and the detected variability of the polarization level in this band is therefore likely not entirely intrinsic to the target. We will therefore exclude the $B$-band results from the interpretation of our results.
We notice that, as visible in Fig. \ref{fig_big}, the measured $B$-band polarization levels do not stand out from other band's results, despite the additional contribution of the Moon. This is however in agreement with our conclusions, since the polarization level induced in the optical by the presence of the Moon should not exceed a few $\%$, that is a comparable level to that due to the emission of synchrotron radiation or to Thomson scattering with free electrons.

\subsection{Interstellar polarization}
Using the average $Q$ and $U$ Stokes parameters of all reference stars on all nights of our monitoring, we could derive the polarization introduced due to interstellar and instrumental effects. Since instrumental polarization is negligible, this value is dominated by interstellar polarization. With the interstellar polarization evaluated in the four bands, we could perform the fit of the data with the Serkowski function, which describes the interstellar polarization expected in our Galaxy \citep{Serkowski75}:

\begin{equation}
\frac{P(\lambda)}{P_{\rm max}}=\rm exp ^{-K\cdot ln^2 (\frac{\lambda_{\rm max}}{\lambda})},
\end{equation}

where $P_{\rm max}$ is the maximum value that the interstellar polarization reaches at a certain wavelength $\lambda_{\rm max}$, $\lambda$ is expressed in $\mu m$, and $K=-0.10+1.86\lambda_{\rm max}$ \citep{Wilking82}.
While $P_{\rm max} $ is typically $<5\%$, $\lambda_{\rm max}$ is normally found between 0.34 and 1 $\mu m$ \citep{Serkowski75}. The fit of our data with this function gives reasonable results, with $\lambda_{\rm max}=(0.44\pm0.08) \mu m$ and $P_{\rm max}=(1.18\pm 0.08)\%$. This value is compatible with the one derived in Sec. \ref{polla_description} (i.e. $P_{\rm max}<1.53\%$) using the simplified version of the Serkowski law.

\subsection{Spectral energy distributions}
Using our LCO and VLT data, broadband spectra could be built. To obtain fluxes from the VLT polarimetric images, we first summed together the o- and e- fluxes measured at each angle of the HWP for the target and a group of reference stars. Then we averaged together the fluxes at the different angles. We then calibrated these fluxes using the XB-NEWS pipeline calibration performed for the LCO telescope images of the same field (see Sec. \ref{lco_description} for details on the XB-NEWS pipeline). The XB-NEWS $i'$ fluxes were converted into $I$-band fluxes using the transformation equations in Tab. 3 of \citet{Jordi2006}. 

\begin{figure*}
\centering
\includegraphics[scale=0.35]{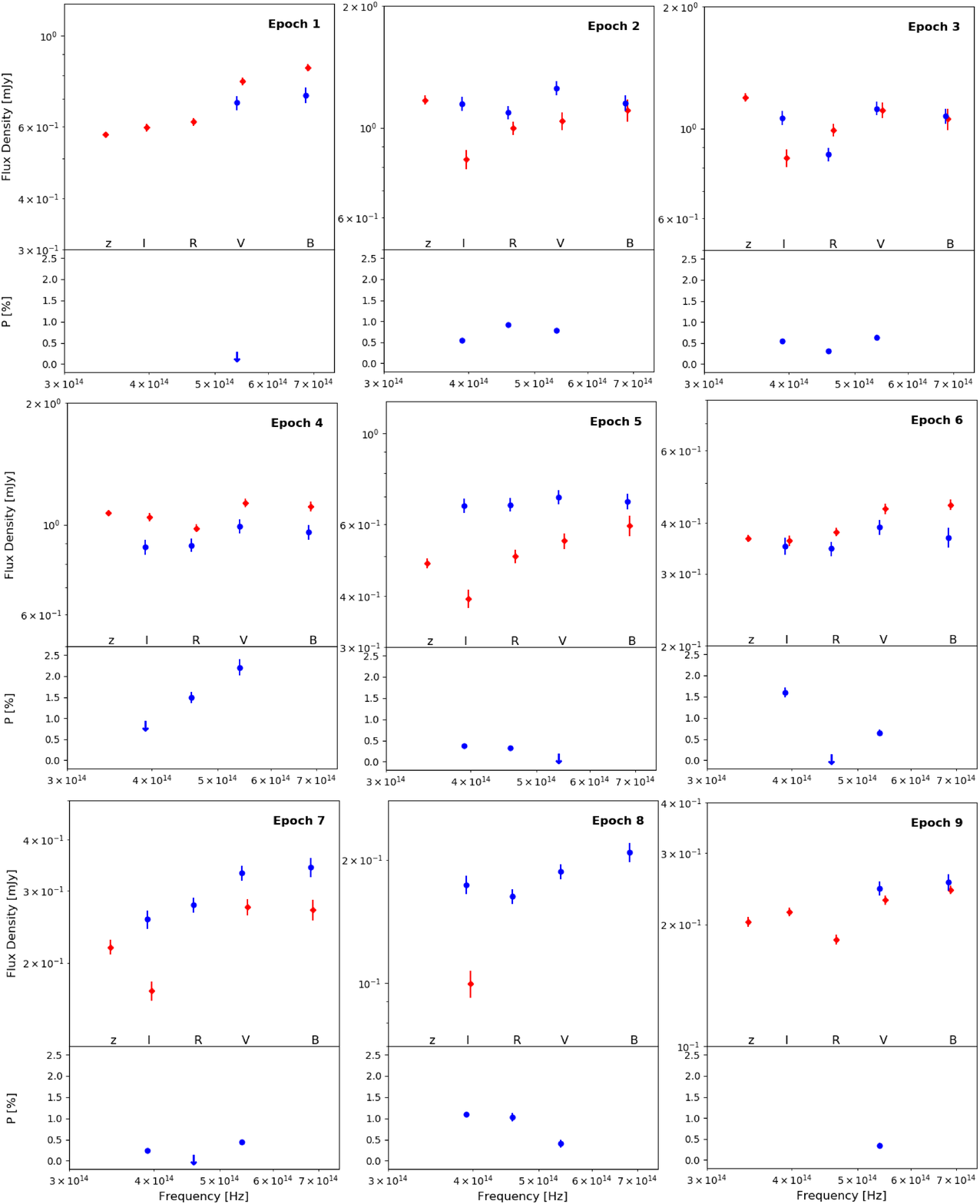}

\caption{\textit{Upper panels}: Broadband spectra of SAX J1808  obtained with Faulkes/LCO (red diamonds) and VLT/FORS2 polarimetric observations (blue circles) during the entire 2019 outburst. All fluxes have been de-reddened using $A_V=0.51\pm 0.04$ mag (from \citealt{DiSalvo2019}) and the \citet{Cardelli1989} relations. A factor of 4 in flux is represented in all plots for comparison purposes. \textit{Lower panels}: Polarization spectra of SAX J1808 (blue circles). Data shown in the figures are reported in Table \ref{polla_results}. 3$\sigma$ upper limits to the LP are represented as blue downward arrows. $B$-band data have been excluded due to moon contamination (see Sec. \ref{Sec:Moon}). We note that the polarization spectra of Epoch 4 might be affected by bad sky conditions (see Table \ref{log_polla}).}
\label{pol_SEDS_tot}
\end{figure*}

To de-redden the VLT and LCO fluxes, we used $A_{V}=0.51\pm 0.04$ \citep{DiSalvo2019}, and the relations of \citet{Cardelli1989} to evaluate the absorption coefficients at all wavelengths. The resulting spectra are reported in Fig. \ref{pol_SEDS_tot}, in comparison with the VLT polarization spectra in all epochs.

We note that the LCO and VLT observations were not taken simultaneously (see Table \ref{log_polla} and \ref{log_LCO}), and therefore the spectra built starting from the two different datasets do not always agree. This is a clear sign of the presence of short time-scale variability at optical frequencies (minutes-hours; see also Fig. \ref{fast_timing_fig}).

\section{The 2015 outburst}\label{Sec_2015}
\begin{figure}
\centering
\includegraphics[scale=0.32]{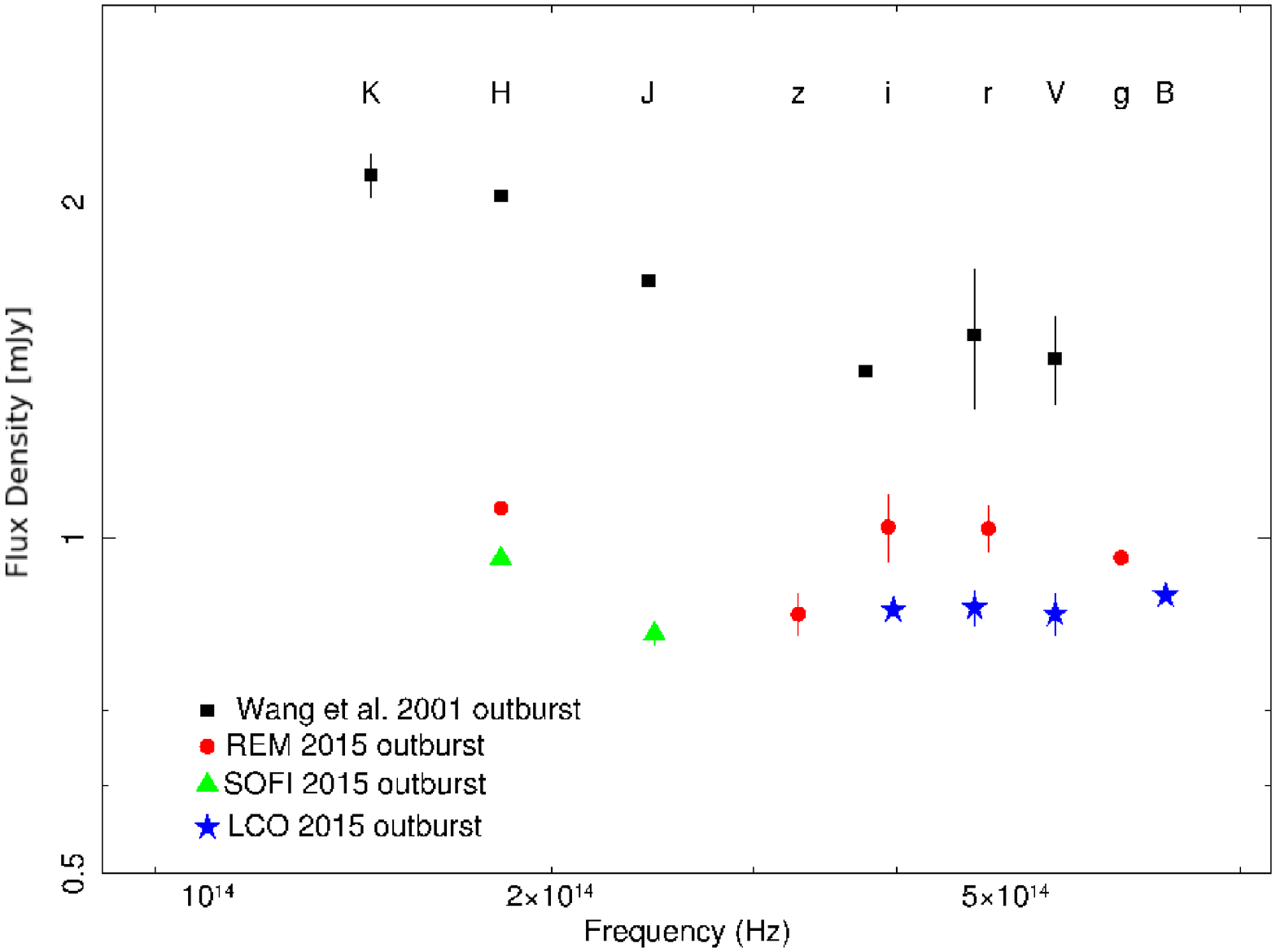}
\caption{Broadband spectrum of SAX J1808 during previous outbursts: with black squares, the 1998 outburst reported in \citet{Wang2001}; with red circles, green triangles and blue stars, the 2015 outburst, with data taken with the REM telescope, the SofI instrument mounted on the NTT, and the LCO network, respectively (Sec. \ref{Sec_2015}).}
\label{SED_prev_outbursts}
\end{figure}

During the 2015 outburst of SAX J1808, we obtained photometric observations in the optical (SDSS $g,r,i,z$ band filters, $400-950\, \rm nm$) and NIR ($H$ band, $\sim 1.6\mu \rm m$) with the Rapid Eye Mount (REM) telescope (La Silla Observatory, Chile) on April 24, 2015 ($\sim$ MJD 57136.2, i.e. during the main outburst decay; see \citealt{Sanna2017} for details on the 2015 outburst evolution). A set of 12 images of 180\,s integration each were acquired simultaneously in the optical bands, together with 15 images of 30\,s integration each in the $H$ band. All images were flat-field and bias corrected using standard procedures, and PSF photometry was performed using the {\tt daophot} tool \citep{Stetson2000}. The flux calibration was performed against the APASS\footnote{\url{https://www.aavso.org/apass-dr10-download}} and the 2MASS\footnote{\url{http://www.ipac.caltech.edu/2mass/}} catalogues. 
SAX J1808 was also observed with LCO as part of the monitoring campaign performed on this source (see Sec. \ref{lco_description} for details). The closest observations to the REM ones were on MJD 57135.8 (April 23).
We could therefore build a NIR-optical broadband spectrum of SAX J1808 (Fig. \ref{SED_prev_outbursts}; we note that the REM spectrum is strictly simultaneous). The spectrum shows the presence of a NIR excess, which is reminiscent of the NIR excess that was detected by \citet{Wang2001} during the 1998 outburst of the source, as shown in Fig. \ref{SED_prev_outbursts} as a comparison. This NIR excess hints at the emission of jets in the system, as previously noted in \citet{Wang2001}. The shape of the NIR excess suggests optically thin synchrotron emission, with a possible jet spectral break in the spectrum around the $J$--$K$ bands, which may extend to the radio regime \citep[see][]{Russell2007,Russell2008AMXPs}.

We also acquired polarimetric observations in the NIR ($J$, $H$ bands) thanks to an accepted ESO DDT proposal (ID:295.D-5012(A); PI: Baglio) at the New Technology Telescope (NTT; La Silla Observatory, Chile) equipped with the SofI instrument. These observations were taken on April 29th 2015, four days after the REM observations (at the very end of the outburst decay, soon before the starting of the reflaring phase; \citealt{Sanna2017}). Due to the absence of a rotating HWP (see Sec. \ref{polla_description}), the entire instrument was rotated at four different positions with respect to the telescope axis ($0^{\circ}, 45^{\circ}, 90^{\circ}, 135^{\circ}$) to obtain the four measurements that are necessary to evaluate the $S$-parameter as described in Sec. \ref{polla_description}. Due to the rotation of the telescope (also the field of view was rotated for different angles), the presence of the polarimetric mask made it challenging to find a large number of reference stars that were present in every image: only five stars could be chosen as reference to correct for instrumental and interstellar effects. We note that, as SofI is mounted at the Nasmyth focus of the NTT, strong instrumental effects can be present, giving up to a few per cent polarization, depending on the direction of the observation. Images were reduced by subtracting the sky background from the frames, and aperture photometry was performed using {\tt daophot}. 
Following the method described in Sec. \ref{polla_description}, we obtained 3$\sigma$ upper limits to the LP in both $J$ and $H$ bands: $P_{J}<1.36 \%$, $P_{H}<1.54\%$.
Summing together the ordinary and extraordinary beams at each angle, we could obtain the flux of SAX J1808 from the SofI observations in $J$ and $H$ bands. As visible in Fig. \ref{SED_prev_outbursts}, the NIR excess is still evident, with fluxes slightly lower with respect to those measured in the REM observations.
We therefore conclude that the jet was present during our observations, as evident from the spectra, and the non-detection of polarization hints at very tangled magnetic fields near the base of the jet. 
This result is consistent with the low optical polarization measured in 2019 by VLT (see below).

\section{Discussion}

In this work we report on the monitoring of the recent outburst of the AMXP SAX J1808.4-3658, using different facilities and modes of observation. In particular, optical ($BVRI$) polarimetric observations have been obtained at nine different epochs during the outburst, between August 8, 2019 and September 25, 2019 (Table \ref{log_polla}). Several detections of intrinsic LP have been found in all bands (see Table \ref{polla_results}), all at a low level ($0.2-2.2\%$). The origin of the LP of the light emitted by the source must be due to a mechanism that is intrinsic to the source, like synchrotron from a jet and/or Thomson scattering in the disc. The interpretation of the results may vary depending on the phase of the outburst in which the target is found during the observations. 

In the broadband spectrum of a LMXB, the optically thin jet spectrum is visualized as a NIR excess with respect to the low-frequency blackbody tail of the accretion disc. The jet spectrum turns over in the infrared, joining the flat or inverted spectrum at lower frequencies in the radio--mm. Typically, jets in BH LMXBs are emitted during the hard state of their outbursts, and are quenched once the source transitions to the soft state. This behaviour has been observed also for NS LMXBs, such as 1RXS J180408.9--342058 \citep{Baglio2016a,Gusinskaia2017}, 4U 1728--34 \citep{Migliari2003} and Aql X--1 \citep{Tudose2009,Miller-Jones2010,DiazTrigo2018}, although the quenching can be less than in BH LMXBs (e.g. \citealt{Migliari2004,Migliari2006}; see also \citealt{Munoz2014}).
SAX J1808 never leaves the hard state during its outbursts, going directly from a hard-state main outburst to a reflaring state (see e.g. \citealt{Patruno2016_1808}). The same happened during the 2019 outburst: according to our LCO monitoring, the optical outburst started on July 30th; the peak of the outburst was reached on $\sim$ August 11th (MJD 58706), and the source entered the reflaring state on August 24th (MJD 58719; \citealt{Baglio2019ATel13103}). Throughout this time, the X-ray spectrum was hard during both the main outburst peak and the reflaring period \citep{Bult2019ATel13001,Bult2019ATel13077,Sanna2019ATel13022}, as typical of faint transients. In particular, a power-law photon index of 1.89 was measured from NuSTAR observations at the outburst peak \citep{Sanna2019ATel13022}, and we find that no significant variations in the hardness ratio can be observed by \textit{Swift}/XRT during the whole duration of the outburst (Hard band: 2--10 keV; Soft band: 0.5--2 keV). Quiescence was then reached at the beginning of October 2019 \citep{Baglio2019ATel13162}.

\begin{figure}
\centering
\includegraphics[scale=0.35]{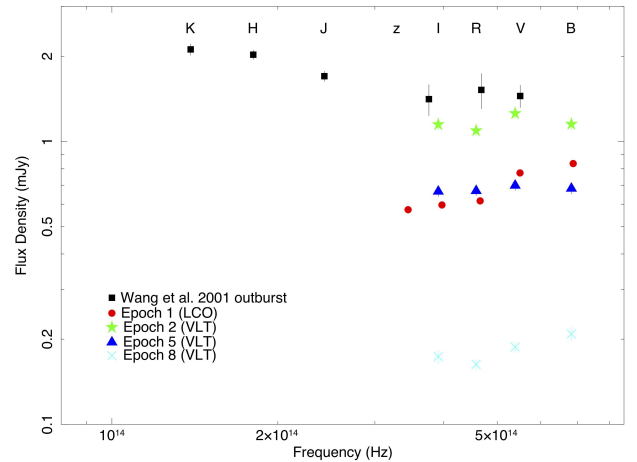}
\caption{Broadband spectrum of SAX J1808 in five different epochs (during the 2019 outburst) for comparison: Epoch 1 during the outburst rise (LCO data), Epoch 2 during the outburst peak, Epoch 5 during the outburst decay and Epoch 8 during the reflaring phase (VLT data). With black squares, the broadband spectrum of the 1998 outburst is reported for comparison \citep{Wang2001}. Only almost simultaneous data have been represented for clarity.}
\label{SED_representative}
\end{figure}

In Fig. \ref{pol_SEDS_tot} we show the comparison between flux and polarization spectra during both the main outburst and the reflaring state phase. A low-frequency ($z$, $I$) excess above the blackbody tail of the accretion disc (that has a positive spectral index) is detected in several epochs (2, 3, 4, 5, 6, 8 and 9; see Fig. \ref{SED_representative} for a selection of epochs), both during the main outburst and the flaring state, suggesting the emission of jets in the system. Epoch 1 instead does not show signatures of jet emission. This is in agreement with radio observations of the target \citep{Williams2019}, according to which no jet was detected on August 4th, but radio emission was observed on August 10th (i.e. between Epoch 1 and Epoch 2). Therefore, the observed optical short-timescale variability after Epoch 1 can be linked to variations in the jet (and/or in the accretion disc). Variability in the jet in particular has been observed in several LMXBs in the past, with rms up to $10-20\%$ in some cases \citep[e.g.][]{Gandhi2010,Gandhi11,Baglio2018}. This is also in agreement with the detected short-timescale variability in the $i'$ band light curve of MJD 58709 shown in Fig. \ref{fast_timing_fig} (left), with a fractional rms of $\sim 3\%$. Being this variability more prominent at the peak of the outburst than at the beginning of the flaring state (MJD 58722; Fig. \ref{fast_timing_fig}, right), it suggests for an origin in a jet that is variable on minutes timescales.

We note that a certain scatter can be observed in the broadband spectra reported in Fig. \ref{pol_SEDS_tot}. This scatter can be explained by considering again the short timescale variability detected in the optical (Fig. \ref{fast_timing_fig}). From the MJD 58709 light curve, we can see that it is possible to observe a variation of $\sim0.15$ mag in $\sim 6$ mins, looking at the lowest and highest flux detections. Therefore, measuring a colour using two filters on a similar timeframe, we could have up to $\sim$0.3 mag difference, which corresponds to a factor of $\sim 1.3$ in flux. Looking at the spectra in Fig. \ref{pol_SEDS_tot}, this variability can therefore explain the observed scatter. Moreover, considering that the LCO and VLT observations on the same Epoch are typically separated in time by a few hours, and taking into account the observed short-timescale variability, a flux change by a factor $\sim 1.5$ can be possible; therefore, also the significant difference between the two datasets in Fig. \ref{pol_SEDS_tot} can be explained.

\begin{figure}
\centering
\includegraphics[scale=0.3, angle=-90]{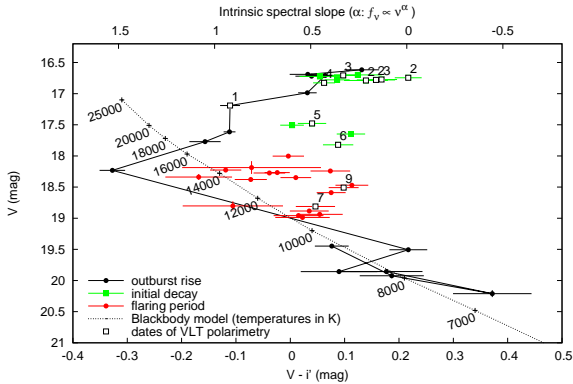}
\caption{Optical color-magnitude diagram of the 2019 outburst of SAX J1808. Bluer colors (greater spectral index) are to the left, redder colors (lower spectral index) are to the right. The three main outburst stages -- the rise, initial fade, and reflaring period are shown by black dots, green squares and red dots, respectively. The blackbody model (see text) is shown by the black dotted line. The rise into outburst is indicated by the black solid lines joining the points. With hollow squares, the dates of VLT polarimetry are represented.}
\label{cmd_fig}
\end{figure}

In Fig. \ref{cmd_fig}, we show the color-magnitude diagram (CMD), $V$ vs. $V-i'$, built using the LCO monitoring data. In the figure, we differentiated between the phases of the outburst (rise, decay, reflares), and we indicated the polarimetric epoch with numbers and squares in the plot. On some dates there were several $V-i'$ paired measurements; each one is shown by the epoch number (short-term variability is evident; the same polarimetric epoch has different positions in the CMD, which means that the relative disc/jet contribution is changing considerably on the timescale of a few hours). With a dotted line, a blackbody model for the accretion disc is represented. This model shows the colour evolution of a single temperature (as opposed to multiple temperatures from different regions), constant area blackbody heating and cooling \citep[for details, see][]{Maitra2008,Russell2011,Zhang2019}. At low temperatures the optical emission originates in the Rayleigh-Jeans tail of the blackbody, and at higher temperatures it originates near the curved peak of the blackbody; this causes the color changes.

We adopt the same method as \cite{Russell2011} to apply this model to the data of SAX J1808. The model assumes an optical extinction, $A_V=0.51\pm0.04$ (see Sec. \ref{polla_description}) to convert $V-i'$ color to intrinsic spectral index (which is indicated in the top axis of Fig. \ref{cmd_fig}). The temperature depends on this color and the normalization of the model depends on several parameters including the size of the accretion disc (which can be estimated from the orbital period and the mass of the neutron star and companion star), the disc filling factor, inclination angle (and potentially, disc warping), and the distance to the source. Due to uncertainties in some of these parameters, and the fact that not all data of SAX J1808 are consistent with this model (see below), we do not fit the model to the data directly, but rather we varied the normalization of the model until it successfully approximated the initial brightening of the source. This part of the outburst is detailed in \cite{Goodwin2020} and it was found that the disc temperature increased from $\sim 7,000$ K to $\sim 10,000$ K during the initial optical brightening, during which hydrogen is ionized and that initiated the fast X-ray and optical rise after MJD $\sim 58700$ \citep{Goodwin2020}.

During the first part of the rise of the main outburst, the $V-i'$ color first decreased as the disc temperature increased to $> 20,000$ K, then the $V-i'$ color increased, becoming redder, diverging from the blackbody model (Fig. \ref{cmd_fig}). After this date the NIR excess is visible in the spectra (Fig. \ref{pol_SEDS_tot}) and the jet component appears to be responsible for the color change in the CMD. The black solid lines in Fig. \ref{cmd_fig} illustrate the evolution of the data in the CMD during the outburst rise to peak.
Epoch 1 is found to be close to the blackbody model, which means that the jet was not contributing considerably at that time (as also visible in the spectrum in Fig. \ref{SED_representative}, where no infrared excess and no significant LP is detected). Epochs 2--4, during the main outburst peak, have the strongest jet contribution and the strongest variability, suggesting high variability in the jet flux (and polarization; see below). Epochs 5 and 6 are during the initial decay and have less jet contribution, but it is still high; Epochs 7--9 instead are during the reflaring state, and a jet contribution might still be present at least in Epoch 7 and 9, but it is less significant than during the main outburst because the data are closer to the blackbody model in the CMD (similarly to Epoch 1). Epoch 8 is not reported in the CMD due to the lack of LCO $V$-band data (Table \ref{log_LCO}). This behaviour is reminiscent of what happened in the case of 1RXS J180408.9-342058 \citep{Baglio2016b}, where the jet was not contributing to the optical LP during the very first stages of its outburst, and was instead observed a few weeks later, during the hard state of the outburst, in the broadband spectra.

Polarization detections have been found during the whole outburst in different epochs and filters (except Epoch 1). As explained in Sec. \ref{Sec:Moon}, we excluded the $B$-band detections from the interpretation of our results, due to the contribution of the Moon, which seems to affect the results in this band. Despite the presence of the low-frequency excess in the spectrum, that is consistent with the emission of optically thin synchrotron radiation (i.e. intrinsically polarized light), the detected LP in all bands is always $\sim 1.6\%$ or less. This very low LP suggests that there are very tangled magnetic field lines in the emitting region, which is near the base of the jet, which give rise to strong cancellation effects and therefore to low levels of intrinsic LP. Since the magnetic field is tangled, and flux variability implies a rapidly changing mass accretion rate that feeds the jet \citep[see e.g.][and references therein]{Gandhi2017}, a certain variability in the LP and angle can also be expected, as we effectively observe, in particular in terms of the LP, but also to some extent in terms of PA. As shown in Fig. \ref{pol_angle_fig}, the PA clusters between $\sim 0^{\circ}$ and $\sim50^{\circ}$, except for two outlier detections in $V$-band during the reflaring state with different polarization angles (epochs 6 and 8).
Excluding $B$-band observations, the PA is clustering in the $\sim 10-50^{\circ}$ range, that is a small range considering the large amount of variability observed in the jet flux. 

\subsection{The main-outburst phase}
\begin{figure*}
\centering
\includegraphics[scale=0.33]{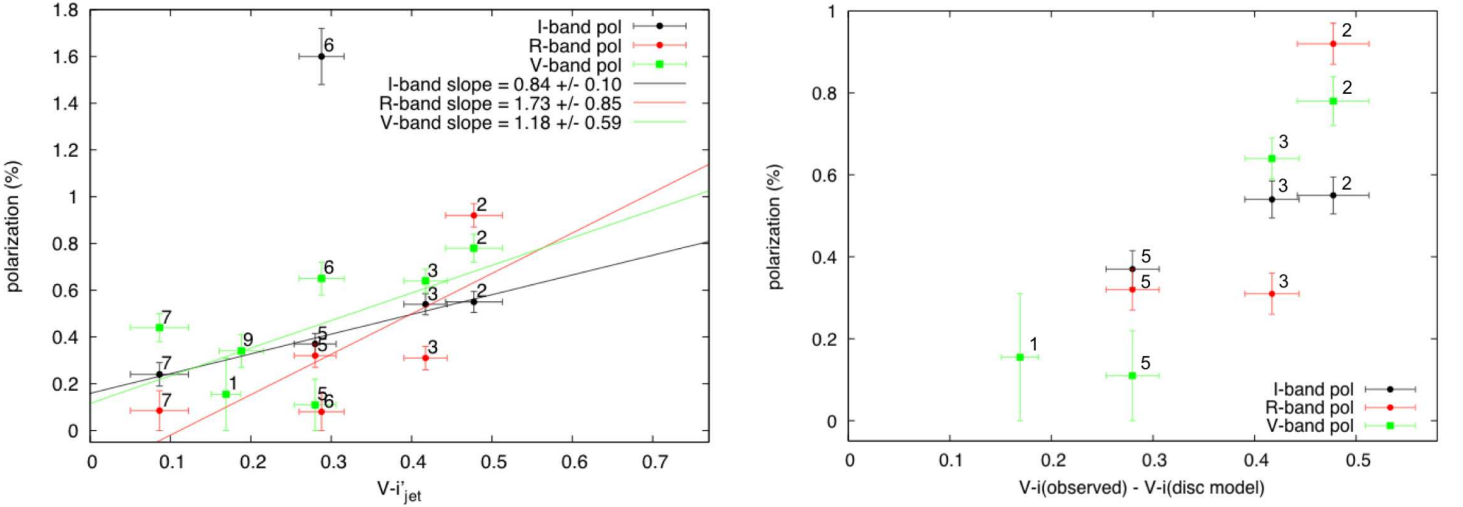}
\caption{\textit{Left panel}: Polarization level vs. $(V-i')_{\rm jet}$ for $V$, $R$, $I$ bands during the whole outburst. Superimposed are linear fits to the data (black, red and green solid lines stand for $I$, $R$, $V$ bands, respectively). Data from Epoch 4 are not reported due to the bad sky conditions of the night. \textit{Right panel}: Polarization level vs. $(V-i')_{\rm jet}$ for $V$, $R$, $I$ bands during the main outburst (Epochs 1, 2, 3, 5). In both panels, epochs are clearly labelled.}
\label{pol_colour_fig}
\end{figure*}

In Fig. \ref{pol_colour_fig} we show a plot of the polarization level in $V$, $R$, $I$ bands vs. the difference between the $V-i'$ observed colour and the $V-i'$ colour due to the disc model, shown as a dotted line in Fig. \ref{cmd_fig} (the data of epoch 4 have been removed from the figure, due to the very bad sky conditions of the night; see Table \ref{log_polla}). The quantity reported on the x-axis (that we will call $(V-i')_{\rm jet}$) shows how far the data are from the disc model, therefore highlighting the contribution of the jet to the total observed flux. When $(V-i')_{\rm jet}<0.2$ mag, the polarization in all bands does not exceed $0.5\%$. This suggests that when the contribution from a jet is lower, the measured LP is also lower. When $(V-i')_{\rm jet}>0.2$ mag instead the jet contribution is stronger, and a variable LP in the $0.3-2\%$ range is observed. Except for a few epochs showing higher LP (polarization flares) in $R$ and $I$ bands, that interestingly all correspond to epochs in the reflaring state, a positive correlation can be found between $(V-i')_{\rm jet}$ and the LP in all three bands. Such a correlation would not be present if the disc (or any high frequency emission component) was causing the observed LP, because it would have made the polarization spectrum bluer.
It seems therefore likely that the red component in the spectrum, i.e. synchrotron emission from the variable jet of SAX J1808, is the cause of the LP, at least during the main outburst phase.

\begin{figure*}
\centering
\includegraphics[scale=0.27]{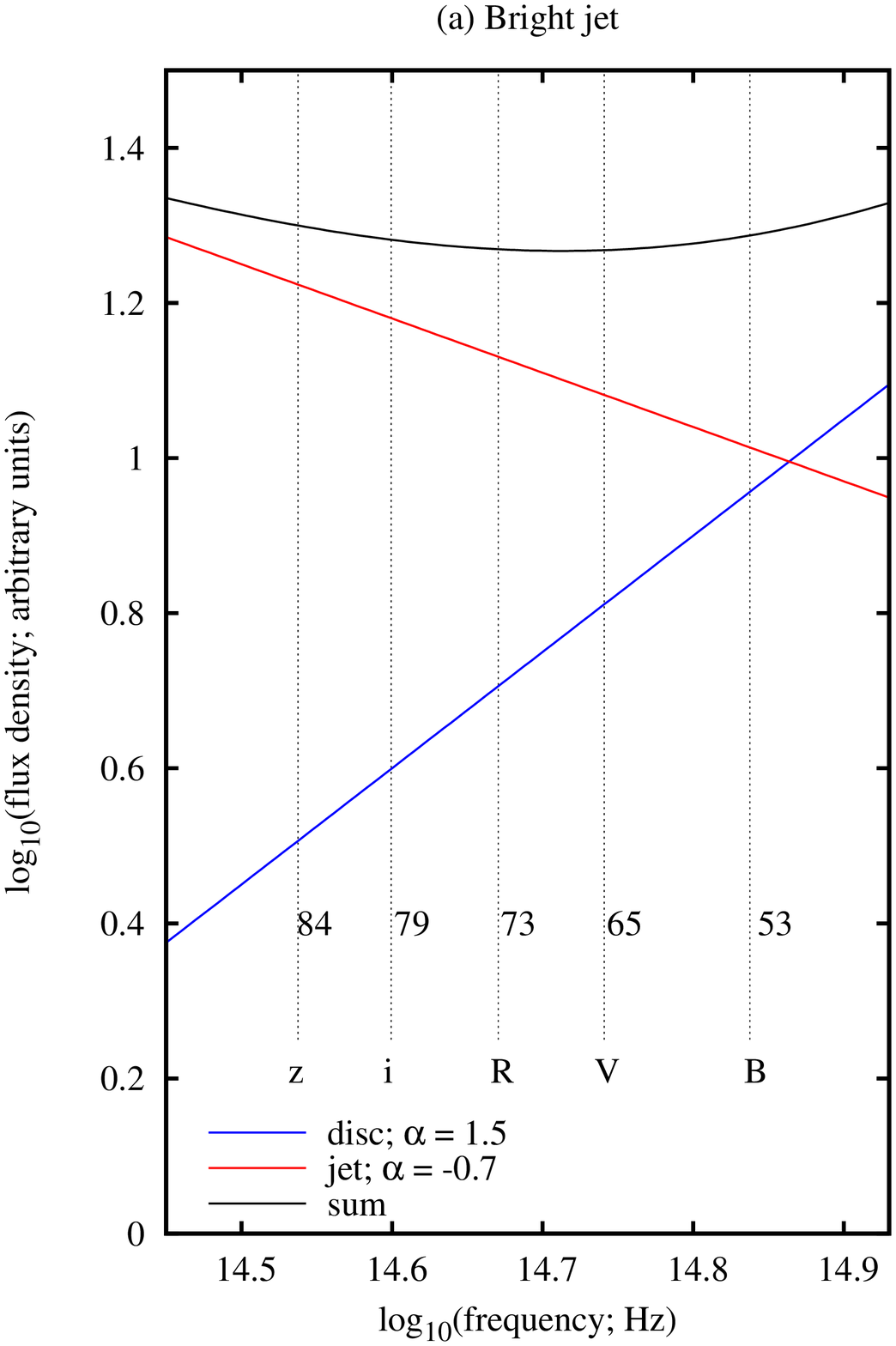}
\includegraphics[scale=0.27]{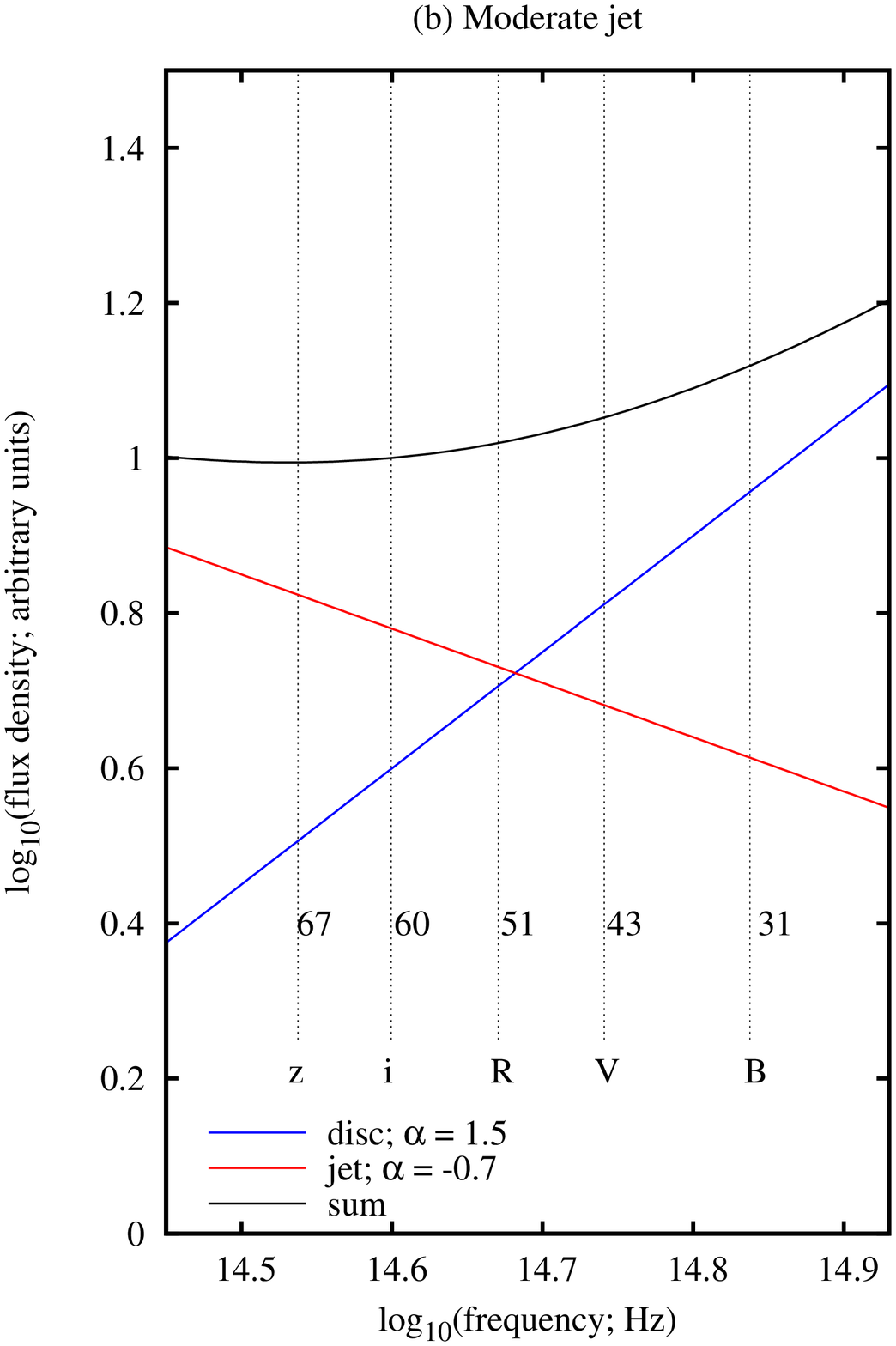}
\includegraphics[scale=0.27]{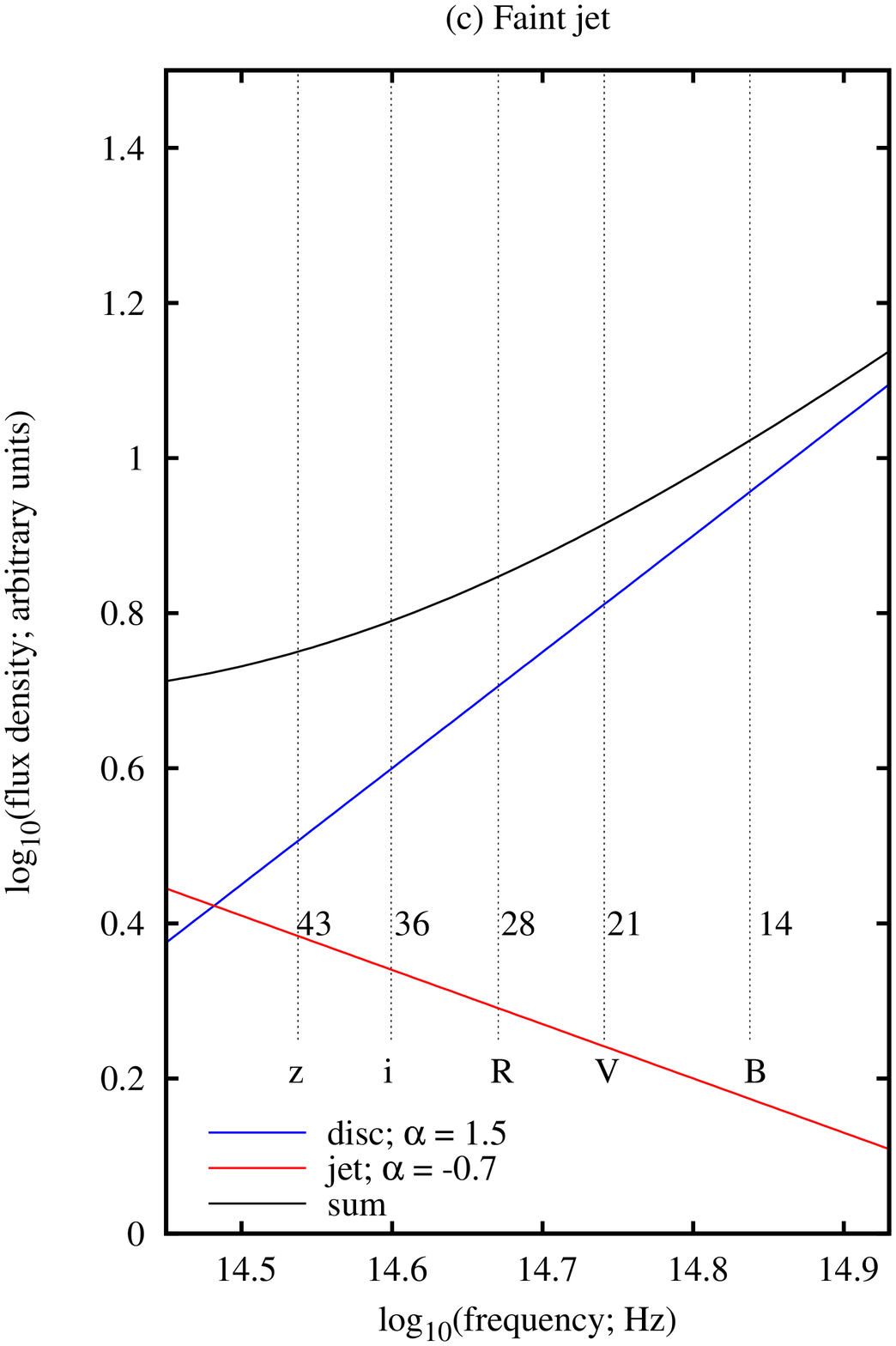}
\caption{Broadband spectrum models for three possible configurations of jet (red line) + disc (blue line) contributions in SAX J1808 during the main outburst phase. The black solid line indicates the sum of the two components. The spectral index of the disc and jet are fixed to 1.5 and -0.7, respectively. With dotted vertical lines the central frequencies of each band are indicated. The jet contribution, as a percentage of the total flux, is indicated in each plot with numbers.  }
\label{sed_models}
\end{figure*}

We notice that this correlation is true in all bands, from $I$ to $V$. Therefore, this conclusion means that the jet polarization is also contributing considerably in $V$-band. This may be surprising, since there is a stronger contribution from the accretion disc at those frequencies. We therefore built some simple broadband spectrum models (Fig. \ref{sed_models}) for a disc (blue line), a jet (red line), and the sum of the two components (black line), approximating the shapes of the spectra to power-laws, since this is a close approximation within this small wavelength range. In particular, according to the color-magnitude diagram shown in Fig. \ref{cmd_fig}, the disc model should have a power-law index of $\sim 1.5$ near the peak of the outburst, whereas the optically thin synchrotron emission from a jet typically has a slope of $\sim -0.7$. Once these have been fixed, we obtained three possible scenarios: 
\begin{itemize}
\item[(a)] the jet is brighter than the disc all the way up to $B$-band. The IR excess is seen in the summed spectrum, getting brighter from $V$ to $z$ bands. The jet contribution, as a percentage of the total flux, is shown as numbers for each filter: the jet produces $84\%$ of the flux in $z$-band, $79\%$ in $i'$, $73\%$ in $R$, $65\%$ in $V$, $53\%$ in $B$. However this is the most extreme case, and the jet is not expected to be that bright in the optical;

\item[(b)] the jet is moderate, the two spectra crossing over around $R$-band. The jet contribution in this case is $67\%$ in $z$-band, $43\%$ in $V$-band. This is a more realistic scenario, and the IR excess is only visible as an increase in flux in the summed spectra for lower frequencies than the $z$-band.

\item[(c)] The jet is faint, producing only $43\%$ of the flux in $z$-band, and $21\%$ in $V$-band.

\end{itemize}

Scenario (b) is the more realistic near the outburst peak; a summed spectrum between scenarios (a) and (b) seems to match the observed spectra in Figs. \ref{pol_SEDS_tot}, \ref{SED_prev_outbursts} and \ref{SED_representative}. For scenario (b) if the jet is producing the polarization, and the jet component is $1\%$ polarized at all optical wavelengths (as an example), then we would expect to see: $0.60\%$ LP in $i'$-band, $0.51\%$ LP in $R$-band, $0.43\%$ LP in $V$-band, $0.31\%$ LP in $B$-band. In particular, we notice that $V$-band LP is $\sim 2/3$ of the $I$-band LP, which shows quantitatively that the jet polarization should be detectable in the $V$-band almost as well as the $I$-band. The relative strength of the polarized signal between different bands depends not just on the polarization of the jet and its variability, but also on the spectral indices and relative flux ratios (see Fig. \ref{sed_models}) of the disc and jet.

If on the other hand, the disc could be producing the observed LP instead of the jet, then for a disc component that is $1\%$ polarized we would obtain: $0.40\%$ LP in $i'$-band, $0.49\%$ LP in $R$-band, $0.57\%$ LP in $V$-band, $0.69\%$ LP in $B$-band. However, in this case we would expect the strongest LP when the data are near the disc model in the color-magnitude diagram (Fig. \ref{cmd_fig}), while the opposite is observed.
We therefore interpret the observed optical polarization during the main outburst of SAX J1808 as induced by the emission of optically thin synchrotron radiation from the compact jet in the system.

We caution the reader that the 3 examples shown in Fig. \ref{sed_models} have been arbitrarily built in order to show representative cases of (a) jet-dominated, (b) moderate jet, and (c) disc-dominated spectra. We can therefore not draw firm conclusions from the derived numbers, but only use them as reference for our study.



\subsection{The reflaring state}

The optical emission processes, and the polarization, could be different during the reflaring state. 
Reflares have been observed at the end of the main outburst for each of the outbursts undergone by SAX J1808 to date \citep{Patruno2016_1808}. Such phenomena have been explained by \citet{Patruno2016_1808} as due to the presence of a strong outflow (caused by a propeller effect) or alternatively of a trapped disc, with limited or absent outflow, in the inner regions of the disc.
We know that jets are present during the reflaring state, from radio emission \citep{Tudor2017} and the IR excess (Figs. \ref{pol_SEDS_tot}, \ref{SED_representative} and \ref{cmd_fig}).

Fig. \ref{pol_SEDS_tot} shows that a low-frequency excess is detected in the last two epochs (Epochs 8 and 9, following the nomenclature reported in Table \ref{log_polla}), as well as on August 24 (Epoch 6 - the first day of the reflaring state/last day of the main outburst phase), suggesting there is emission of optically thin synchrotron from a jet also during the reflaring state. Polarization detections are also reported in this phase. The colour-magnitude diagram in Fig. \ref{cmd_fig} also shows that a contribution from the jet is still present during the flaring state (red points in the figure), however it is less prominent than during the main outburst phase. Moreover, Fig. \ref{pol_colour_fig} (left panel) shows that a positive correlation between the measured LP in all bands and the quantity $(V-i')_{\rm jet}$ is present both during the main outburst and the reflaring state, despite the presence of three polarization flares in $I$ and $R$ bands, all happening during the reflaring period.
Therefore, similar conclusions as in the main outburst phase can be drawn, with the observed LP being due to the emission of a jet in the system. 
The observed $I$- and $R$-band polarization flares on Epochs 6 and 8 (Fig. \ref{pol_colour_fig}, left panel) could be another manifestation of the high variability of the jet of SAX J1808, also inferred from the colour magnitude diagram in Fig. \ref{cmd_fig}. In particular, since the polarization flares are observed at low-frequency only, it is likely that the component that is responsible for them has a red spectrum, and so is due to the component that produces the IR excess. 
In addition, on August 24th (Epoch 6), we only have an $I$-band polarization flare, with LP of $1.6\%$, whereas in $R$-band the LP on the same epoch is very low ($<0.16\%$), and the $R$-band observations were performed $\sim 22$ mins after the $I$-band ones. This means that the polarization flare at this epoch was happening on a timescale of a few tens of minutes. On September 7th (Epoch 8) instead, the polarization flare is observed in both $I$- and $R$-band, with LP of $1-1.1\%$, and a lower $V$-band LP of $\sim 0.4\%$. This is suggestive for a red spectrum (i.e. the jet), and of a possible longer duration with respect to the flare observed in Epoch 6 (it might last longer than the full set of observations, i.e. $>75$ min).  

An alternative and intriguing interpretation to the LP that is observed during the reflaring state is that it could be linked to Thomson scattering with the propelled matter. The propeller scenario is in fact used as one explanation of the reflaring activity that is observed in SAX J1808 \citep{Patruno2016_1808}, implying the presence of matter, also in the form of free electrons, that is expelled in the form of a wide angle outflow, and can therefore scatter the light emitted by the accretion disc, giving rise to LP. The presence of propelled matter is also supported by the detection of optical and UV millisecond pulsations during the 2019 outburst, as reported in Ambrosino, Miraval Zanon et al. (submitted).
This scenario has been invoked to explain the variable LP measured for PSR J1023+0038 by \citet{Hakala2018}. However, if this was the case for SAX J1808, the spectrum of its polarization flares would mimic the spectrum of the disc, being therefore bluer.
Since the presence of a jet during the flaring state of SAX J1808 is proven by radio observations and by the infrared excess in the spectra, we could also consider the possibility that the propelled matter might Thomson-scatter the light emitted by the jet (instead of that emitted by the accretion disc). This could give us as a result a red polarization spectrum of the reflares, similarly to what we observe. However, the propelled matter typically generates a wide angle outflow, and therefore it is unlikely that the radiation from the jet, that is collimated, can interact with it causing Thomson scattering to happen (and therefore LP).

In general, the observed polarization flares do not fit the main trend shown in the plot in Fig. \ref{pol_colour_fig}. Our observations are suggestive of a red spectrum also for the polarization flares (which would go against the interpretation of the observed LP being due to Thomson scattering of disc photons in the propeller outflow), but it is difficult to be conclusive since we only have a few such LP flares, and their spectrum on short timescales is not well constrained. If the jet is responsible for the polarization flares, then it is possible that its strong variability (supported by Fig. \ref{cmd_fig} and by the variability of the spectra seen in Fig. \ref{pol_SEDS_tot}) is associated with a variation in the level of ordering of the magnetic field lines in the jet.
A similar behaviour has been observed e.g. in the case of the NS LMXB Sco X-1 and the BHs LMXB GX 339--4 and V404 Cyg \citep{Russell08,Russell2011pol,Shahbaz2016,Lipunov2016,Lipunov2019}, where strong LP variability in the NIR or optical was observed, and was interpreted in terms of variability in the jet flux and/or in the conditions in the inner regions of the jet.

\subsection{Magnetic field ordering}

Since the LP is interpreted as due to synchrotron from the jet, the level of ordering of the magnetic field lines $f$ can be estimated following \citet{Bjornsson1982} (see also eq. 7 of \citealt{Russell14}). In particular:

\begin{equation}\label{eq_ordering}
    LP(\%)=100\,f\, \frac{1-\alpha_{\rm thin}}{5/3 - \alpha_{\rm thin}},
\end{equation}

where $f$ goes from 0 (non-uniform, tangled magnetic field lines) to 1 (uniform and aligned) and $\alpha_{\rm thin}=-0.7\pm 0.2$ is the typical value of the optically thin synchrotron spectrum for a LMXB.

From scenario (b) of Fig. \ref{sed_models}, we see that the jet is likely contributing about $\sim 60\%$ of the total flux emitted by SAX J1808 in $i'$ band. Considering our highest polarization detection in $I$-band (1.6$\%$ in Epoch 6), this translates into a LP of the radiation emitted by the jet only in $I$-band of $(2.7\pm 0.2)\%$. Using eq. \ref{eq_ordering}, we therefore get $f=0.04\pm0.01$.  
Our weakest detection instead (i.e. $0.24\% \pm 0.05\%$ in Epoch 7) translates into an LP of $(0.4\pm0.08)\%$ for the jet only, from which we get $f=0.01\pm0.01$.
This level of ordering is lower with respect to other LMXBs (see e.g. the LMXB GRO J1655-40 for which $f= 0.41\pm0.19$; \citealt{Russell08}). However, the LP measured for SAX J1808 is quite similar to the LP measured in the NIR for (e.g.) GX 339-4 for a jet contributing at the $100\%$ to the total emission of the source, and also to the LP of the NS LMXB Sco X-1 \citep{Russell08}; we therefore conclude that the low level of $f$ measured for SAX J1808 in this work is in line with what could be inferred for other LMXBs.

A similar result can also be obtained for the 2015 outburst (Sec. \ref{Sec_2015}). We performed a fit of the optical spectrum obtained with REM with an irradiated blackbody, and we extrapolated it to the $H$-band frequency, resulting in a flux density of the blackbody of 0.3 mJy in $H$-band. From the comparison of this result with the measured flux density of SAX J1808 at the same frequency (1.1 mJy), we estimate that the jet is contributing $\sim 70\%$ to the total emission of the system. Therefore, the upper limit to the LP in $H$-band that we reported in Sec. \ref{Sec_2015} translates into an upper limit of 2.14$\%$ for the light emitted from the jet only. From eq. \ref{eq_ordering} we could therefore estimate the maximum level of ordering of the magnetic field, $f<0.03$, which is very low but comparable with what we obtained for the detections of LP during the 2019 outburst.

\section{Conclusions}
In this work we have presented the results of a photometric and polarimetric optical campaign performed during the 2019 outburst of the AMXP SAX J1808.4-3658.
The source has been found to be significantly linearly polarized in several epochs and filters, with low polarization ($<2.2\%$, reaching down to $\sim 0.2\%$ in one epoch). Moreover, a flux excess which is stronger at longer wavelengths was observed in the spectrum of the target in most epochs, which is likely due to optically thin synchrotron emission from jets. From the analysis of the $V-i'$ colour compared to a blackbody disc model, we found that in general during the main outburst, when the contribution of the jet was higher (i.e. the spectrum is redder), a higher level of LP was observed. The results suggest that the observed LP during the main outburst is caused by synchrotron from a jet, contributing from $I$-band up to the $V$-band, with variable polarization. During the reflaring phase, there is less jet contribution and in general, less polarization, except some flares of polarization, with evidence suggesting they are happening on timescales of tens of minutes. The spectrum of the polarization is still red, which is suggestive of a jet origin. Other scenarios (like Thomson scattering with free electrons in the disc or in the propelled matter) can be ruled out both during the main outburst and the reflaring phase by the red spectrum of the observed polarization.
We therefore conclude that the optical LP of SAX J1808 is linked to the launching of jets; in particular, the low value of the polarization in all epochs and bands is a signature of strongly tangled magnetic fields near the base of the jet, with a level of ordering derived from $I$-band observations of 1--4$\%$.

Finally, the measured polarization angle clusters around a value of PA $\sim 30 \pm 20^{\circ}$ during the whole campaign, implying the magnetic field vector has an angle of $\sim 120 \pm 20^{\circ}$. In other NS and BH LMXBs, the magnetic field has been found in most cases to be parallel to the jet axis in this optically thin synchrotron region \citep[][and references therein]{Russell2018pol,Shahbaz2019Review}, which suggests that the jet in SAX J1808 may be launched at a position angle of $\sim 120 \pm 20^{\circ}$. This could be tested with future high resolution VLBI radio observations.

In addition we presented a short campaign performed on SAX J1808 during the 2015 outburst, with the emission of jets being detected as a NIR excess in the spectral energy distribution during the decay of the outburst; however, no LP was observed in the NIR ($P<1.5\%$). This can be a sign of very tangled magnetic fields at the base of the jet and/or of high variability in the jet component, as observed in the 2019 outburst. In particular, the estimated upper limit to the magnetic field ordering during the 2015 outburst is of the $3\%$, in line with what derived for the 2019 outburst.

The polarimetric results reported here support the jet being the origin of the NIR excess commonly observed in AMXPs, and demonstrate that optical and infrared polarimetry can constrain the magnetic field structure, and its variability, close to the launching region of jets in NS LMXBs. If such observations, including polarimetric variability studies on minute timescales, can be performed on other NS LMXBs, this will go someway to uncovering the general magnetic field properties of jets in LMXBs, which can be compared to that of AGN and other jet launching sources.

\section*{Acknowledgements}
We thank the anonymous referee for useful comments and
suggestions.
Based on observations collected at the European Southern Observatory under ESO programmes 0103.D-0575 and 295.D-5012.
The Faulkes Telescope Project is an education partner of Las Cumbres Observatory (LCO). The Faulkes Telescopes are maintained and operated by LCO. We acknowledge the support of the NYU Abu Dhabi Research Enhancement Fund under grant RE124.
NM acknowledges financial support from ASI-INAF contract No. 2017-14-H.0.
TMD acknowledges support via the Spanish research grants RYC-2015-18148 and AYA2017-83216-P.
PDA, SCa, SCo acknowledge support from ASI grant I/004/11/3.

\bibliographystyle{aasjournal}



\end{document}